\documentclass[twocolumn,prb,aps,superscriptaddress]{revtex4}
\usepackage[latin9]{inputenc}
\usepackage{amsmath}
\usepackage{amssymb}
\usepackage{graphicx}
\usepackage{color}

\newcommand{\mathsym}[1]{{}}
\newcommand{\unicode}[1]{{}}

\makeatletter

\usepackage{dcolumn}
\usepackage{bm}

\usepackage{amsmath}
\usepackage{amssymb}
\usepackage{graphicx}
\usepackage{slashbox}



\begin{document}

\title{Theory of the evolution of magnetic order in Fe$_{1+y}$Te compounds
with increasing interstitial iron }

\author{Samuel Ducatman}

\affiliation{Department of Physics, University of Wisconsin-Madison, Madison,
WI 53706, USA}
\affiliation{School of Physics and Astronomy, University of Minnesota, Minneapolis,
MN 55455, USA}

\author{Rafael M. Fernandes}

\affiliation{School of Physics and Astronomy, University of Minnesota, Minneapolis,
MN 55455, USA}

\author{Natalia B. Perkins}

\affiliation{Department of Physics, University of Wisconsin-Madison, Madison,
WI 53706, USA}
\affiliation{School of Physics and Astronomy, University of Minnesota, Minneapolis,
MN 55455, USA}

\begin{abstract}
We examine the influence of the excess of interstitial Fe on the magnetic
properties of Fe$_{1+y}$Te compounds.
Because in  iron chalcogenides the correlations are stronger
than in the iron arsenides, we assume in our model that some of the
Fe orbitals give rise to localized magnetic moments. These moments
interact with each other via exchange interactions as well as phonon-mediated
biquadratic interactions that favor a collinear double-stripe state,
corresponding to the ordering vectors $\left(\pm\pi/2,\pm\pi/2\right)$.
The remaining Fe orbitals are assumed to be itinerant, giving rise
to the first-principle derived Fermi surface displaying nesting features
at momenta $\left(\pi,0\right)/\left(0,\pi\right)$. Increasing the
amount of itinerant electrons due to excess Fe, $y$, leads to changes
in the Fermi surface and to the suppression of its nesting properties.
As a result, due to the Hund's coupling between the itinerant and
localized moments, increasing $y$ leads to modifications in the local
moments' exchange interactions via the multi-orbital generalization
of the long-range Ruderman-Kittel-Kasuya-Yosida (RKKY) interaction.
By numerically computing the RKKY corrections and minimizing the resulting
effective exchange Hamiltonian, we find, in general, that the excess
electrons introduced in the system change the classical magnetic ground
state from a double-stripe state to an incommensurate spiral, consistent
with the experimental observations. We show that these results can
be understood as a result of the suppression of magnetic spectral weight of the
itinerant electrons at momenta $\left(\pi,0\right)/\left(0,\pi\right)$,
combined with the transfer of broad magnetic spectral weight from
large to small momenta, promoted by the introduction of excess Fe.
\end{abstract}
\maketitle

\section{Introduction}

Fe$_{1+y}$Te chalcogenides are the parent compounds of the simplest
family of iron-based superconductors.~\cite{bao09,xia09,liu10,Lipscombe11,rodriguez11,stock11,zaliznyak12,mizuguchi12,roessler11,koz12,koz13,parshall12,rodriguez13}
Both the electronic and magnetic properties of Fe$_{1+y}$Te compounds
show strong sensitivity to the amount of non-stoichiometric Fe ions.
For small values of $y$, the low-temperature crystal structure is
monoclinic ($P21/m$), and the magnetic order is described by the
commensurate propagation vector ${\mathbf{Q}}=(\pm\pi/2,\pm\pi/2)$,
corresponding to a double-stripe pattern. This is remarkably different
from FeAs-based parent compounds, which display an antiferromagnetic
order described by ${\mathbf{Q}}=(\pi,0)$ or ${\mathbf{Q}}=(0,\pi)$,
corresponding to
 single-stripe patterns, and a crystal structure with orthorhombic
symmetry ($Pmmn$). Experimentally, it is observed that by increasing
the amount of interstitial Fe, the magnetic structure of Fe$_{1+y}$Te
becomes an incommensurate spiral. The incommensurate ordering manifests
itself as a shift in the elastic neutron scattering peak with respect
to the ${\mathbf{Q}}=(\pm\pi/2,\pm\pi/2)$ positions. According to
neutron scattering experiments,\cite{bao09,rodriguez11,stock11,zaliznyak12,parshall12,rodriguez13}
in the range of $0.11<y<0.16$ the shift is approximately along the
diagonal directions ${\mathbf{Q}}=(\pi/2-\delta,\pi/2-\delta)$ or
${\mathbf{Q}}=(-\pi/2+\delta,\pi/2-\delta)$; however, $\delta$ does
not vary smoothly with $y$.

There have been several theoretical attempts to understand the magnetic
properties of Fe$_{1+y}$Te compounds within the localized spin scenario,
\cite{Ma09,fang09,ducatman12,chen13,JPHu12} since these materials
are known to be more strongly correlated than their arsenide counterparts.
\cite{Kotliar11} Although the magnetic order at both low and at high
levels of Fe excess can be successfully described by a $J_{1}-J_{2}-J_{3}$
super-exchange model, it is clear that the local picture alone cannot
describe the magnetic properties of Fe$_{1+y}$Te, as it requires
$y$-dependent exchange couplings. Alternatively, this property indicates
that itinerant electrons are also important to describe the magnetism
of these materials, suggesting that hybrid models with coupled localized-itinerant
moments are a suitable starting point. \cite{Mazin09,Kruger09,yin10,Lv10,Liang12,Dagotto12,imada13}

In this paper, we argue that the evolution of the magnetic interactions
due to $y$-dependent charge doping is the key to understand the experimentally
observed magnetic phase diagram of Fe$_{1+y}$Te. Our study is based
on the assumption that, in these particular iron chalcogenides, some
of the Fe orbitals are almost localized while the other orbitals remain
itinerant. This idea is supported by recent dynamical mean-field theory
studies of FeTe systems~\cite{haule09,lanata13} showing that the
Hund's coupling can promote an orbital-selective localization already
in the paramagnetic phase.\cite{Bascones12} Here, we demonstrate
that the change in magnetic properties observed in the Fe$_{1+y}$Te
compounds can be reasonably well captured by an effective model in
which localized spins
 acquire a long-range RKKY-type interaction,~\cite{akbari11,akbari13,Andersen14}
in addition to the $J_{1}-J_{2}-J_{3}$ Heisenberg super-exchange\cite{Ma09}
and  biquadratic couplings. \cite{wisocki11,mazin14,Paul11_PRL,Fernandes10,paul11}
We note that in Refs.\cite{wisocki11,mazin14}  the biquadratic term was calculated
on a purely electronic basis, and obtained model gave a good
agreement with experimentally measured spin-wave spectrum in several Fe-based materials.
  Another possible origin  of the biquadratic terms is due to the magnetoelastic coupling.\cite{Paul11_PRL,Fernandes10,paul11}

The former, mediated by the multi-orbital spin susceptibility of the
itinerant electrons, is sensitive to the addition of excess Fe $y$,
rendering the magnetic ground state of the local spins
 change as a function of $y$. In particular, we find that the shift
in the chemical potential promoted by the excess electrons changes
the Fermi surface in a way that suppresses the $\left(\pi,0\right)/\left(0,\pi\right)$
peaks of the itinerant spin susceptibility of the parent compound,
 promoting at the same time peaks at small-momentum values. As a result,
the effective exchange interaction $J_{2}$ is suppressed, $J_{3}$
is enhanced, and $J_{1}$ changes sign, favoring a spiral incommensurate
state, in contrast to the double-stripe state of the stoichiometric
compound.

The outline of the paper is as follows. In Sec. II, we present
an effective super-exchange model describing localized magnetic moments
in the multi-band correlated electron sea and argue that this is
 a suitable minimal microscopic model to describe the magnetism of
Fe$_{1+y}$Te compounds. In Sec. III, we study the evolution of
the Fermi surface of FeTe$_{1+y}$ with increasing level of the Fe
excess using the tight-binding model (TBM) originally  proposed by  Ma \textit{et al}\cite{Ma09}, which had been later also used by  Wang \textit{et al}\cite{wang10} to explain further resutls in FeTe$_{1+y}$.
We show that while the Fermi surface at small $y$ has both small
hole pockets at the $\Gamma$- and $M$-points and elliptical electron
pockets at the $X$- and $Y$-points, at large $y$ all pockets are
electron-like. In Sec. IV, we study spin fluctuations in the correlated
multi-band electron system and compute the Pauli susceptibility within
the random phase approximation (RPA). We find that at small $y$ the
itinerant spin susceptibility peaks at $(\pi,0)$ and $(0,\pi)$ due
to the weak nesting between the hole and the electron pockets
 connected by these momenta. The $y$-dependent charge doping suppresses
the $(\pi,0)$ and $(0,\pi)$ peaks, but it leads to an increase of
the spin fluctuation in the central part of the Brillouin zone. In
Sec. V, we compute the RKKY interactions. We first perform a qualitative
computation of the RKKY interactions using a simple phenomenological
model, and then perform a quantitative analysis based on the realistic
RPA susceptibility obtained in Sec. IV. In Sec. VI, the classical
phase diagram of the effective spin model is presented. In agreement
with experimental findings, the computed phase diagram displays a
transition, above a certain level of Fe excess, from a commensurate
double-stripe phase, characterized by the wave-vectors ${\bf Q}=(\pi/2,\pm\pi/2)$
or ${\bf Q}=(\pm\pi/2,\pi/2)$, to an incommensurate spiral (IC) phase
characterized by the wave-vector ${\bf Q}=(q,q)$. We conclude with
a summary in Sec. VII. The paper has two appendices. Appendix A contains
the derivation of the biquadratic exchange couplings arising from
the magneto-elastic coupling. Appendix B provides the explicit expression
of the classical energy of the effective super-exchange model describing
FeTe$_{1+y}$.

\begin{figure*}
\label{fig1} \includegraphics[width=0.65\columnwidth]{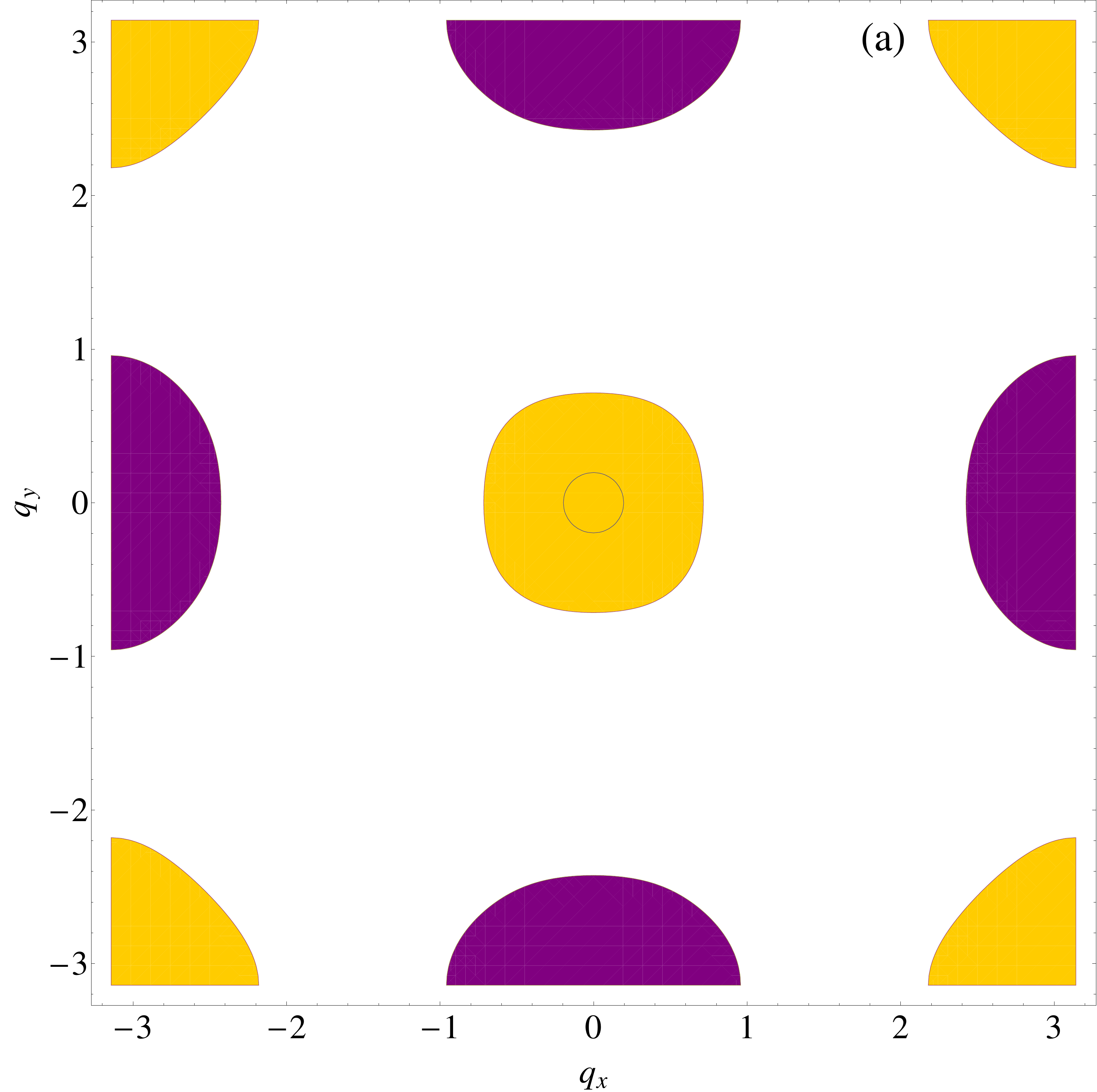} \includegraphics[width=0.65\columnwidth]{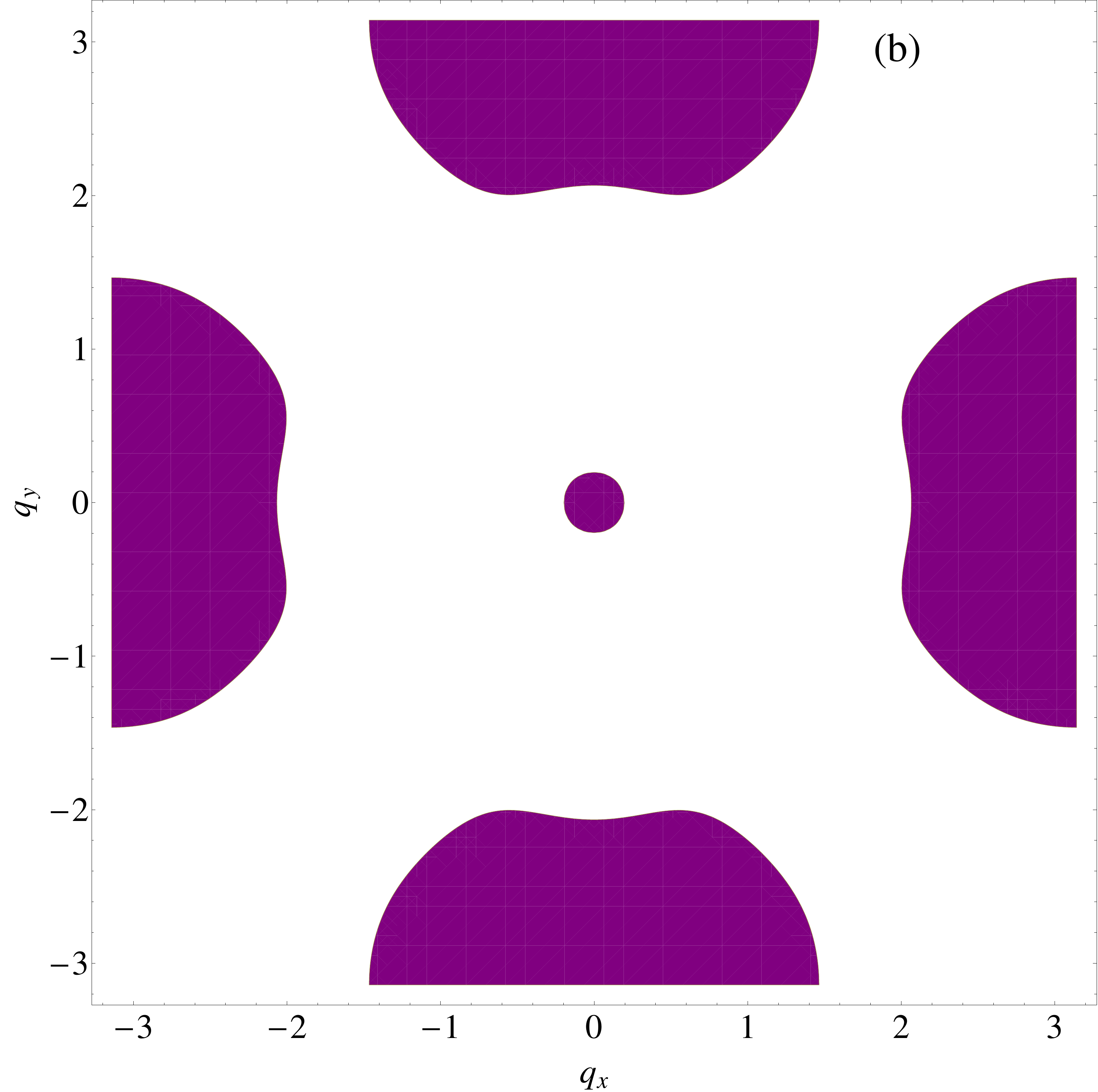}
\includegraphics[width=0.65\columnwidth]{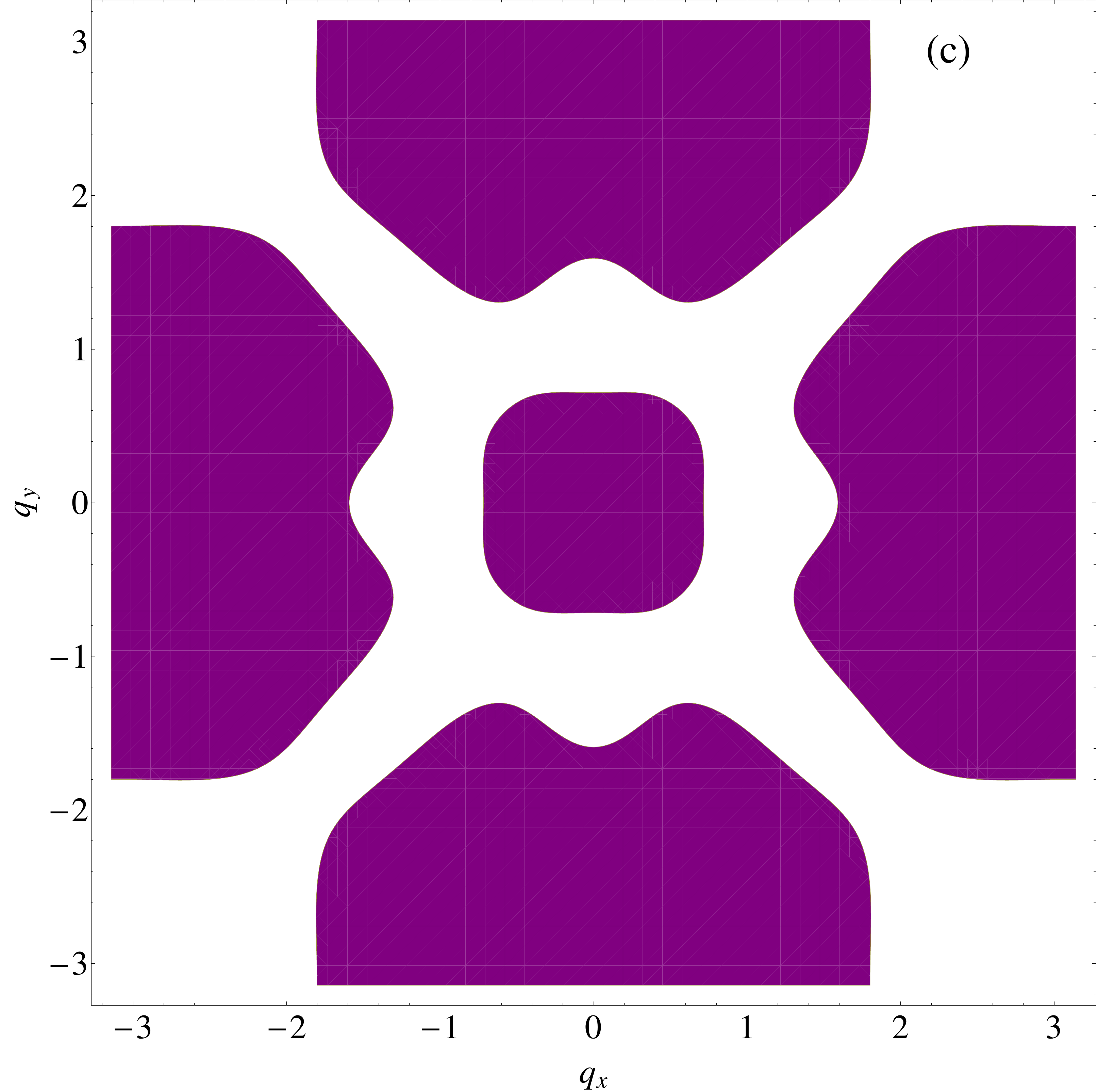} \caption{Fermi surfaces of the TBM describing
the itinerant electrons obtained for different values of the chemical
potential $\mu$: (a) $\mu=0$ eV, (b) $\mu=0.4$ eV, (c) $\mu=0.8$
eV. The hole-like and the electron-like pockets are shown by yellow
 (light) and purple (dark) regions, correspondingly. The primary contribution
to the electron pockets is from the $xz$ and $yz$ orbitals, whereas
 the hole pocket at the corner of the Brillouin zone is mainly of
$xy$ character. }
\end{figure*}

\section{The Model}

To capture the fact that in Fe$_{1+y}$Te correlations lead to different
levels of itineracy in distinct Fe orbitals,~\cite{haule09,lanata13}
we consider a semi-phenomenological ``hybrid'' model containing both localized
and itinerant moments -- similar in spirit to the models of Refs.
\cite{yin10,Lv10,Liang12}. In particular, we assume that the electrons
occupying the $x^{2}-y^{2}$ and $3z^{2}-r^{2}$ orbitals form local
moments with $S=1$ due to the Hund's coupling.
The remaining $3d$
electrons have itinerant character and can be controlled by the concentration
of excess Fe as $y$ increases.
A full derivation of such an effective model from the microscopic Hamiltonian is an intricate problem beyond the scope of this work.

 In most of the paper, we assume that
each excess Fe atom contributes eight electrons,~\cite{Savrasov09} but
our main conclusions do not change if one considers that less electrons
are introduced by each Fe.\cite{Singh10,ding13} To account for the
changes in the Fermi surface promoted by these excess Fe atoms, we
employ a rigid band approximation in which the doping of excess electrons
shifts the chemical potential from $\mu=0$ to positive values.

Thus, the microscopic Hamiltonian we use is an effective double-exchange
model describing localized magnetic moments in the multi-band correlated
electron sea, which can be written as
\begin{eqnarray}
H=H_{e}+H_{\mathbf{S}}+H_{{\mathbf{\sigma}}\mathbf{S}}~.\label{ham1}
\end{eqnarray}

The first term describes the interacting itinerant electrons:
\begin{eqnarray}
H_{e}=H_{0}+H_{int}.\label{ham2}
\end{eqnarray}

The non-interacting part $H_{0}$ is based on the five-orbital TBM
\begin{eqnarray}
H_{0}=\sum_{\mathbf{k},a,b,\sigma}\left(t_{\mathbf{k}\sigma}^{ab}c_{\mathbf{k}a\sigma}^{\dagger}c_{\mathbf{k}b\sigma}+
H.C.\right)~,\label{H0}
\end{eqnarray}
 where $c_{\mathbf{k}\sigma a}^{\dagger}$ denotes the creation operator
for an electron of momentum $\mathbf{k}$ with spin $\sigma$ in the
orbital $a$, and $t_{\mathbf{k}\sigma}^{ab}$ are the tight-binding
matrix elements. Here, we are interested only in the contribution coming
from the $xz$, $yz$, and $xy$ orbitals, as explained above. Since
the other orbitals do not contribute to the Fermi surface, the static spin susceptibility, which will give rise to the RKKY interactions, is very similar regardless of whether we consider a projected three-orbital model or the actual five-orbital tight-binding model.

The interaction part consists of four terms:\cite{Kubo07,graser09,Zhang09}
\begin{align}
H_{int}= & U\sum_{i,a}c_{ia\uparrow}^{\dagger}c_{ia\uparrow}c_{ia\downarrow}^{\dagger}c_{ia\downarrow}\nonumber \\
 & +U'\sum_{i,a\neq b,\sigma,\sigma'}c_{ia\sigma}^{\dagger}c_{ia\sigma}c_{ib\sigma'}^{\dagger}c_{ib\sigma'}\nonumber \\
 & +J_{H}\sum_{i,a\neq b,\sigma,\sigma'}c_{ia\sigma}^{\dagger}c_{ib\sigma'}^{\dagger}c_{ia\sigma'}c_{ib\sigma}\nonumber \\
 & +J'\sum_{i,a\neq b}c_{ia\uparrow}^{\dagger}c_{ia\downarrow}^{\dagger}c_{ib\downarrow}c_{ib\uparrow}\label{H_int}
\end{align}
 where $U$ is the intra-orbital Coulomb repulsion, $U'$ is the inter-orbital
Coulomb repulsion, $J_{H}$ is the Hund's coupling, and $J'$ is the
pair-hopping term. Hereafter, we set $U'=U-2J_{H}$ and $J'=J_{H}$
to ensure the invariance of the Hamiltonian under rotations in orbital
space.

The second term in Eq.(\ref{ham1}) describes the interaction between
the localized spins:
\begin{eqnarray}
H_{\mathbf{S}}=\sum_{ij}\left(J_{ij}\mathbf{S}_{i}\cdot\mathbf{S}_{j}-\frac{K_{ij}}{S^{2}}(\mathbf{S}_{i}\cdot\mathbf{S}_{j})^{2}\right),\label{hamloc}
\end{eqnarray}
 where $J_{1},J_{2}$, and $J_{3}$ are super-exchange couplings between
first-, second-, and third-nearest neighbors. In this work, we use
the values of $J_{1},J_{2}$, and $J_{3}$ obtained from the first-principles
electronic-structure calculations.~\cite{Ma09} $K_{ij}$ denote
generalized non-Heisenberg exchange couplings between first and second
neighbors and ring-exchange. These couplings are predominantly determined
by the magneto-elastic coupling, and their derivations are shown in
Appendix A.
In particular, here we consider the first- and second-neighbors biquadratic
couplings $K_{1}$ and $K_{2}$, as well as a ``diagonal'' ring-exchange
$K_{\mathrm{diag}}=-K_{2}$.

The third term
\begin{eqnarray}
H_{\mathbf{\sigma}\mathbf{S}}=J_{H}\sum_{j,a}\mathbf{\boldsymbol{\sigma}}_{ja}\cdot\mathbf{S}_{j}\label{hamjh}
\end{eqnarray}
 describes the coupling between the localized spins, $\mathbf{S}_{j}$,
and the itinerant electrons $\mathbf{\boldsymbol{\sigma}}_{ja}$.
Here, $\mathbf{S}_{j}$ are the localized spins of electrons on orbitals
$x^{2}-y^{2}$ and $3z^{2}-r^{2}$, and $\mathbf{\sigma}_{ja}$ are
the spins of itinerant electrons on orbitals $a=xy,yz,xz$ defined
as
\begin{eqnarray}
\boldsymbol{\mathbf{\sigma}}_{ja}=\frac{1}{N}\sum_{\mathbf{k},\mathbf{k'},\sigma,\sigma^{\prime}}e^{i(\mathbf{k'}-\mathbf{k})\cdot\mathbf{R_{j}}}c_{\mathbf{k}a\sigma}^{\dagger}\mathbf{\boldsymbol{\tau}}_{\sigma\sigma^{\prime}}c_{\mathbf{k'}a\sigma^{\prime}}~,
\end{eqnarray}
where $\mathbf{\boldsymbol{\tau}}_{\sigma\sigma^{\prime}}$ are the
Pauli matrices and
 $N$ is the number of lattice sites. For our investigations, the
sign of $J_{H}$ is not important, as shown in the following. Due to the interaction
described by Eq.(\ref{hamjh}), the itinerant electrons mediate additional
exchange couplings between the localized moments:
\begin{eqnarray}
H_{\mathrm{RKKY}}=\sum_{ij}J_{ij}^{\mathrm{RKKY}}\mathbf{S}_{i}\cdot\mathbf{S}_{j}.\label{RKKY}
\end{eqnarray}

The couplings $J_{ij}^{\mathrm{RKKY}}$ between localized spins on
lattice sites $\mathbf{R}_{i}$ and $\mathbf{R}_{j}$, known as RKKY
interactions, can be obtained by integrating out the itinerant degrees
of freedom. In particular, $J_{ij}^{\mathrm{RKKY}}$ are determined
by the static spin susceptibility $\chi({\mathbf{q}},\omega=0)=\chi({\mathbf{q}})$
of the multi-band conduction electron sea:
\begin{eqnarray}
J_{ij}^{\mathrm{RKKY}}= & - & J_{H}^{2}\chi(\mathbf{R}_{i}-\mathbf{R}_{j})=\label{RKKYcouling}\\
 & - & J_{H}^{2}\sum_{\mathbf{q}}e^{\imath(\mathbf{R}_{i}-\mathbf{R}_{j})\cdot{\mathbf{q}}}\chi({\mathbf{q}}).\nonumber
\end{eqnarray}

Taking into account both Heisenberg and RKKY interactions, the effective
low-energy Hamiltonian can then be written as
\begin{eqnarray}
H_{\text{eff}}=\sum_{ij}J_{ij}^{\text{eff}}\mathbf{S}_{i}\mathbf{S}_{j}-\sum_{ij}\frac{K_{ij}}{S^{2}}(\mathbf{S}_{i}\cdot\mathbf{S}_{j})^{2},\label{exchange-model}
\end{eqnarray}
 where, for convenience, we defined $J_{ij}^{\text{eff}}=J_{ij}+J_{ij}^{\mathrm{RKKY}}$.

\section{Fermi surface evolution}

In this section, we discuss how the Fermi surface of Fe$_{1+y}$Te evolves
with increasing amount of interstitial Fe, $y$.   To compute the Fermi surface, we use
the TBM, those matrix elements  we present in Table I.

\begin{widetext}
\begin{center}
\begin{table}
\caption{Tight-binding hopping matrix elements of the TBM.\cite{Ma09,wang10}
All $t_{ij}(\mathbf{R})$, where $i,j$ are orbital indexes and $\mathbf{R}$ are distances between Fe ions, are given in eV.  We use the following notations: $\mathbf{R}=(x,y)$, where $x=l_x a_x$, $y=l_y a_y$  and $\mathbf{a}$ is the unit vector of the one-Fe unit cell.
 Thus, (1,0) column corresponds to the hopping to the nearest neighbor in the $x$ direction, (1,1) column  corresponds to the hopping between second neighbors along the diagonal, etc. Last three columns
 define how the hopping elements change when one applies inversion symmetry (I), $C_4$ rotation which switches
 $x\rightarrow y$ and mirror plane symmetry which changes $y \rightarrow -y$, respectively.
 Also, the relation $t_{ij}(\mathbf{R})=t_{ji}(-\mathbf{R})$ stands.
  For shortness,   we denote orbital indexes we as $1=3z^2-r^2$, $2=xz$, $3=yz$, $4=xy$, $5=x^2-y^2$.
 The uniform energy shifts of the diagonal elements $t_{ii}$ (not shown in the Table)  are given by $\epsilon_1=-0.449$, $\epsilon_2=0.111$, $\epsilon_3=0.111$, $\epsilon_4=-0.077$, and $\epsilon_5=-0.366$ (in eV).
  } 

\label{T1}
\begin{tabular}{|c|c|c|c|c|c|c|c|c|}\hline
  \backslashbox{$t_{ij}$}{$\mathbf{R}$(x,y)}  & (1,0)  & (1,1) & (2,0) & (2,1) & (2,2) & I & $x\rightarrow y$ & $y\rightarrow -y$
\\\tableline

$t_{11}$ & $0.0164 $ & $-0.033$ &   $-0.0131$ & $0      $ & $-0.0154$ & $t_{11}$ & $t_{11}$ & $t_{11}$ \\\tableline
$t_{12}$ & $-0.126$ & $0     $ &   $-0.0125$ & $0      $ & $0      $ & $-t_{12}$ & $-t_{12}$ & $-t_{13}$\\\tableline
$t_{13}$ & $0.126 $ & $-0.206$ &   $0.0125 $ & $-0.0178$ & $-0.0262$ & $-t_{13}$ & $t_{13}$ & $-t_{12}$\\\tableline
$t_{14}$ & $0      $ & $0.0894$ &   $0      $ & $0      $ & $-0.0112$ & $t_{14}$ & $t_{14}$ & $-t_{14}$     \\\tableline 
$t_{15}$ & $-0.356 $ & $0     $ &   $-0.0301$ & $-0.0102$ & $0      $ & $t_{15}$ & $-t_{15}$ & $t_{15}$ \\\tableline
$t_{22}$ & $-0.217 $ & $0.131 $ &   $-0.0178$ & $0.0132 $ & $-0.0119$ & $t_{22}$ & $t_{22}$ & $t_{33}$ \\\tableline
$t_{23}$ & $0.12   $ & $0     $ &   $0.0326 $ & $-0.0283$ & $0      $ & $t_{23}$ & $-t_{23}$ & $t_{23}$ \\\tableline
$t_{24}$ & $0.207  $ & $0     $ &   $0      $ & $0      $ & $0      $ & $-t_{24}$ & $-t_{24}$ & $t_{34}$ \\\tableline
$t_{25}$ & $-0.302 $ & $0.143 $ &   $0      $ & $0      $ & $0      $ & $-t_{25}$ & $t_{25}$ &  $-t_{35}$\\\tableline
$t_{33}$ & $-0.217 $ & $0.376 $ &   $-0.0178$ & $-0.0394$ & $0.0839 $ & $t_{33}$ & $t_{33}$ & $t_{22}$ \\\tableline
$t_{34}$ & $0.207  $ & $0.115$ &   $0      $ & $0.0221 $ & $-0.0129$ & $-t_{34}$ & $t_{34}$ & $t_{24}$\\\tableline
$t_{35}$ & $0.302  $ & $0     $ &   $0      $ & $0.0349 $ & $0      $ & $-t_{35}$ & $-t_{35}$ & $-t_{25}$\\\tableline
$t_{44}$ & $0.0305 $ & $0.0904$ &   $0.0103 $ & $-0.0181$ & $-0.0292$ & $t_{44}$ & $t_{44}$ & $t_{44}$\\\tableline
$t_{45}$ & $0      $ & $0     $ &   $0      $ & $0.0145 $ & $0      $ & $t_{45}$ & $-t_{45}$ & $-t_{45}$\\\tableline
$t_{55}$ & $0.397  $ & $-0.0508$&   $-0.0448$ & $0      $ & $0.0213 $ & $t_{55}$ & $t_{55}$ & $t_{55}$\\\hline
\end{tabular}
\end{table}
\end{center}
\end{widetext}

\begin{figure}
\label{fig2} \includegraphics[width=0.8\columnwidth]{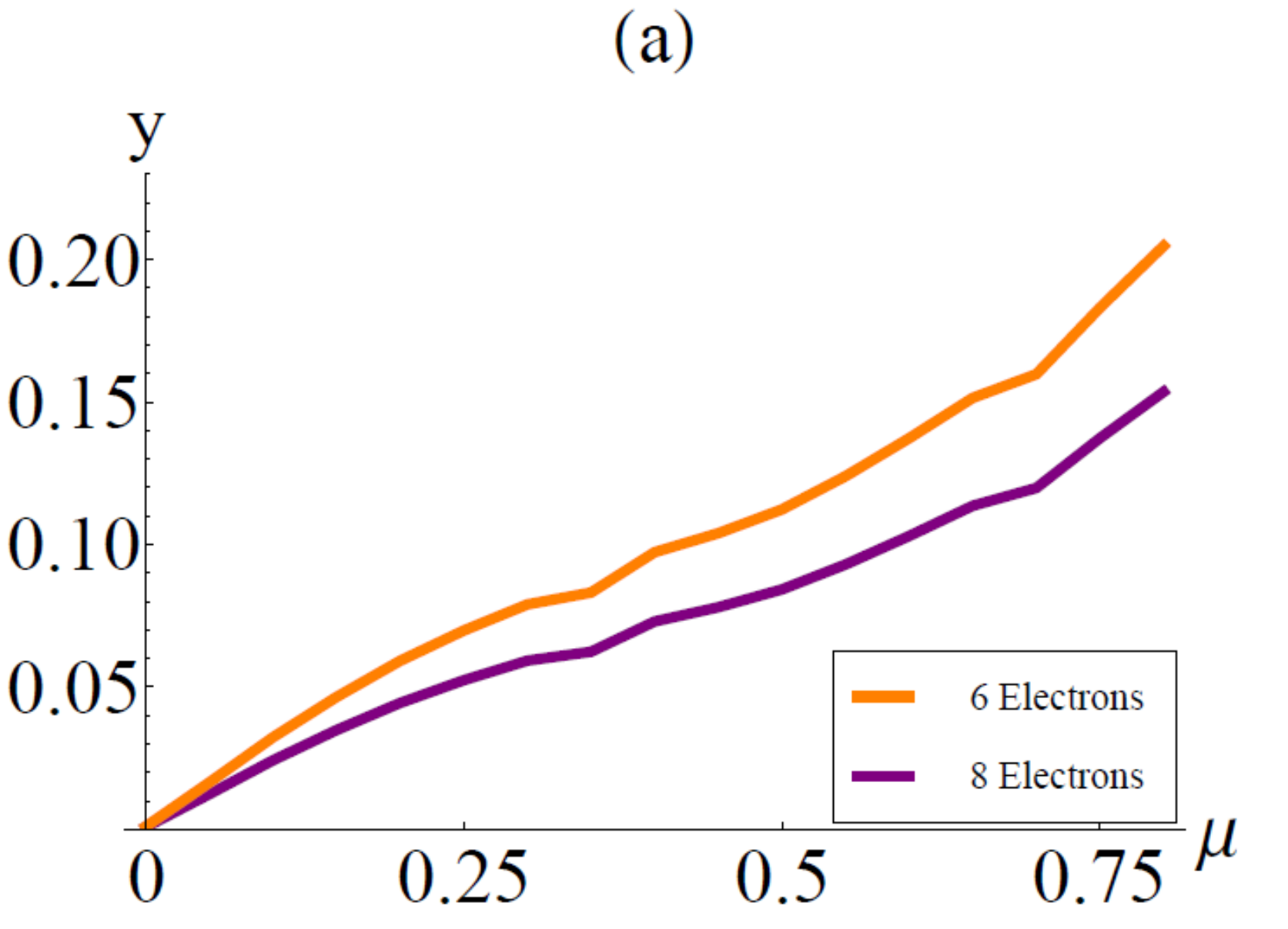}
\includegraphics[width=0.8\columnwidth]{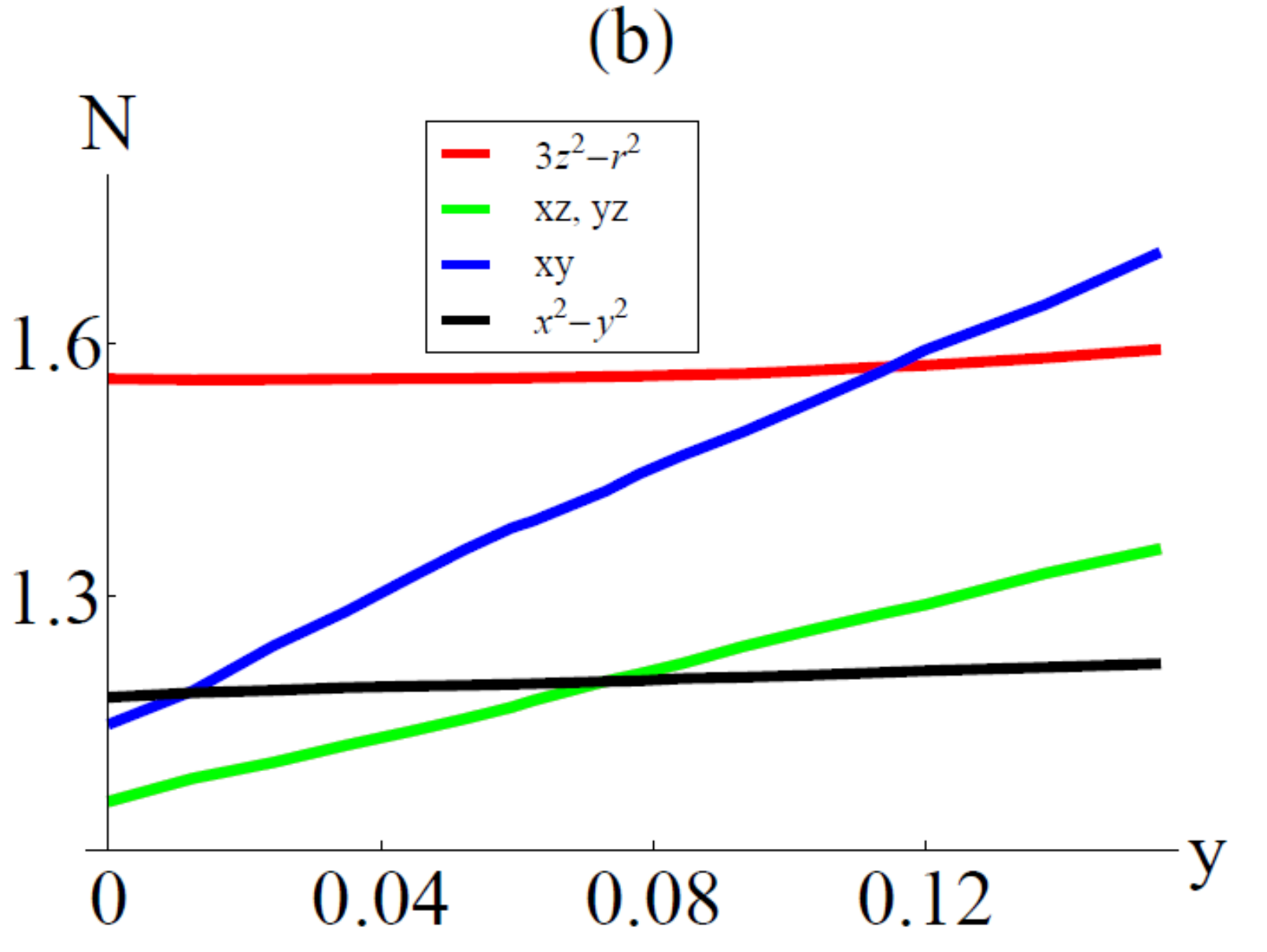}
\caption{(a) The dependence of the chemical potential
shift $\mu$ on the concentration of excess iron $y$ when each excess
iron provides 6 electrons (orange) and 8 electrons (purple). (b) The
orbitally resolved electron occupation numbers $N$ as functions of
$y$. The occupation of $3z^{2}-r^{2}$, $xz$, $yz$, $xy$, and
$x^{2}-y^{2}$ orbitals are shown by red, green, brown, blue, and
black lines, respectively. The $xz$ and $yz$ orbitals have the same
electron occupation numbers due to the tetragonal symmetry of the
system.}
\end{figure}

In Fig. 1 (a), we
show the Fermi surface of FeTe ($y=0$) obtained from the
TBM. Since
the pockets have predominantly $xz$, $yz$, and $xy$ character,
we present the Fermi surface without projecting out the localized-spin
orbitals $x^{2}-y^{2}$ and $3z^{2}-r^{2}$. Similar to most iron
pnictide parent compounds, the Fermi surface is characterized by elliptical
electron pockets at the $X=(\pi,0)$ and $Y=(0,\pi)$ points and circular
hole pockets centered at the $\Gamma=(0,0)$ and $M=(\pi,\pi)$ points.
We note a weak nesting between electron and hole pockets
connected by the wave vectors $(\pi,0)$ and $(0,\pi)$, but no Fermi
surface nesting associated with the magnetic ordering vector ($\pm\pi/2,\pm\pi/2$).
This observation is in agreement with an angle-resolved photoemission
(ARPES) study of the topology of the Fermi surface,\cite{xia09} which
also did not observe nesting corresponding to the magnetic ordering
vector.

When the level of interstitial Fe increases, the chemical potential
shifts up, and the geometry of the Fermi surface changes significantly.
In Figs. 1 (b) and (c) we plot constant energy cuts which correspond
to the chemical potential shifts (b) $\mu=$0.4 eV and (c) $\mu=$0.8
eV. The nesting between electron and hole pockets disappears very
quickly as the sizes of the hole and electron pockets change significantly
with the shift of the chemical potential $\mu$. Initially, while
the size of the electron pockets is enlarged, the size of the hole
pockets is reduced. The hole pockets disappear completely at $\mu=0.23$
eV. Then, at bigger shifts of $\mu$, the pocket
 at the $\Gamma$  point changes its character and becomes electron-like
at $\mu=0.3$ eV. This general behavior is also in agreement with
ARPES measurements in electron-doped iron arsenides.\cite{Liu11}

In order to relate the shift of the chemical potential $\mu$ to the
amount of excess iron $y$, in Fig.2 (a) we plot the dependence of
$\mu$ on $y$ for two cases,
within the rigid band approximation: in the first case, each interstitial
Fe adds eight electrons (purple line) and in the second case~\cite{Savrasov09},
each excess Fe atom has the same valence as the non-interstitial atoms,
adding six electrons (orange line).\cite{Singh10,ding13} Of course,
the general trends are the same, the differences being in the specific
values of $y$ correspondent to each chemical potential shift. For
instance, a shift of 0.4 eV (Fig. 1 (b)) corresponds to $y=0.07$
and $y=0.09$ , if we consider that each interstitial Fe adds eight and
six electrons, respectively, whereas the shift of 0.8 eV (Fig. 1 (c))
corresponds to $y=0.15$ and $y=0.20$, respectively, for 8 and 6
electrons. Hereafter, all results are computed assuming that each
interstitial iron adds 8 electrons into the band.

In Fig. 2 (b), we show how the electron occupation numbers for different
orbitals depend on the
concentration of excess Fe, $y$. We see that the occupations of the
$xz$, $yz$, and $xy$ orbitals change significantly, indicating that
the charge doping due to the Fe excess goes predominantly to these
orbitals. The occupations of the $x^{2}-y^{2}$ and $3z^{2}-r^{2}$
orbitals barely change with an increase of $y$, which is consistent
with the fact that these orbitals do not contribute to the Fermi surface.
Notice, that due to tetragonal symmetry, the occupation numbers of
$xz$ and $yz$ orbitals are exactly the same and the corresponding
lines completely overlap.

\section{Spin susceptibility of itinerant electrons}
\label{ss}

Having established how the low-energy itinerant states change as a function
of excess Fe, we now discuss in details the form of the multi-orbital
spin susceptibility $\chi({\mathbf{q}},\omega)$ as function of $y$.
Here we use the formalism which was originally developed for the 
five-orbital model for the Fe-pnictides in Ref.\cite{graser09} and later
extensively studied for various orbital models in Ref.\cite{brydon11}.

In the paramagnetic state, the spin-rotation invariance requires that
transverse and longitudinal components of the spin susceptibility
are identical. Thus, we can express $\chi({\mathbf{q}},\omega)$ only
in terms of the components of the transverse susceptibility: $\chi({\mathbf{q}},\omega)=\frac{3}{2}\chi^{+-}({\mathbf{q}},\omega)$.
We note that as we are dealing with multi-orbital systems, the spin
susceptibility is a four-index tensor, while the total susceptibility
is a sum over all components of this tensor.

\begin{figure*}
\label{fig3} \includegraphics[width=0.65\columnwidth]{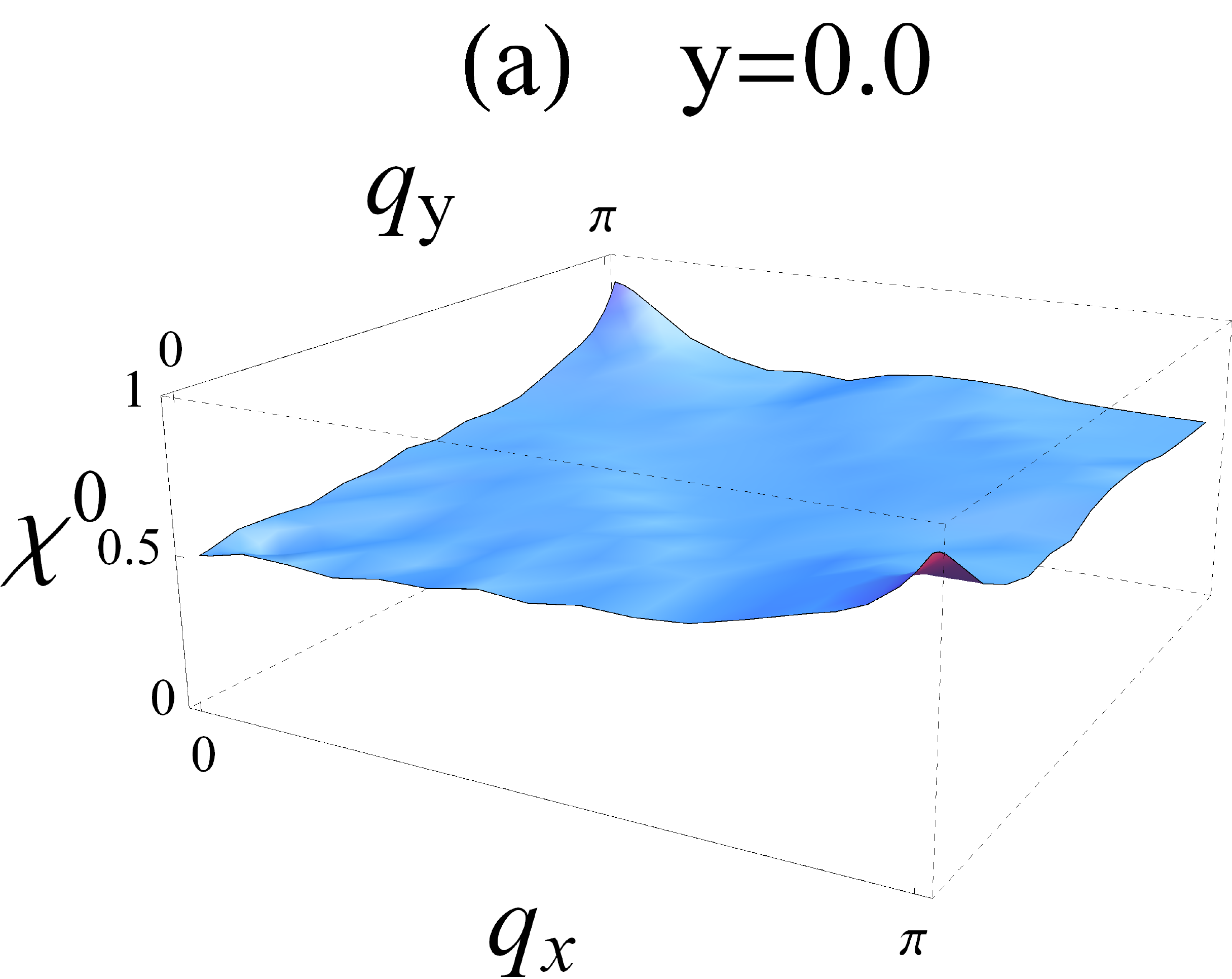} \includegraphics[width=0.65\columnwidth]{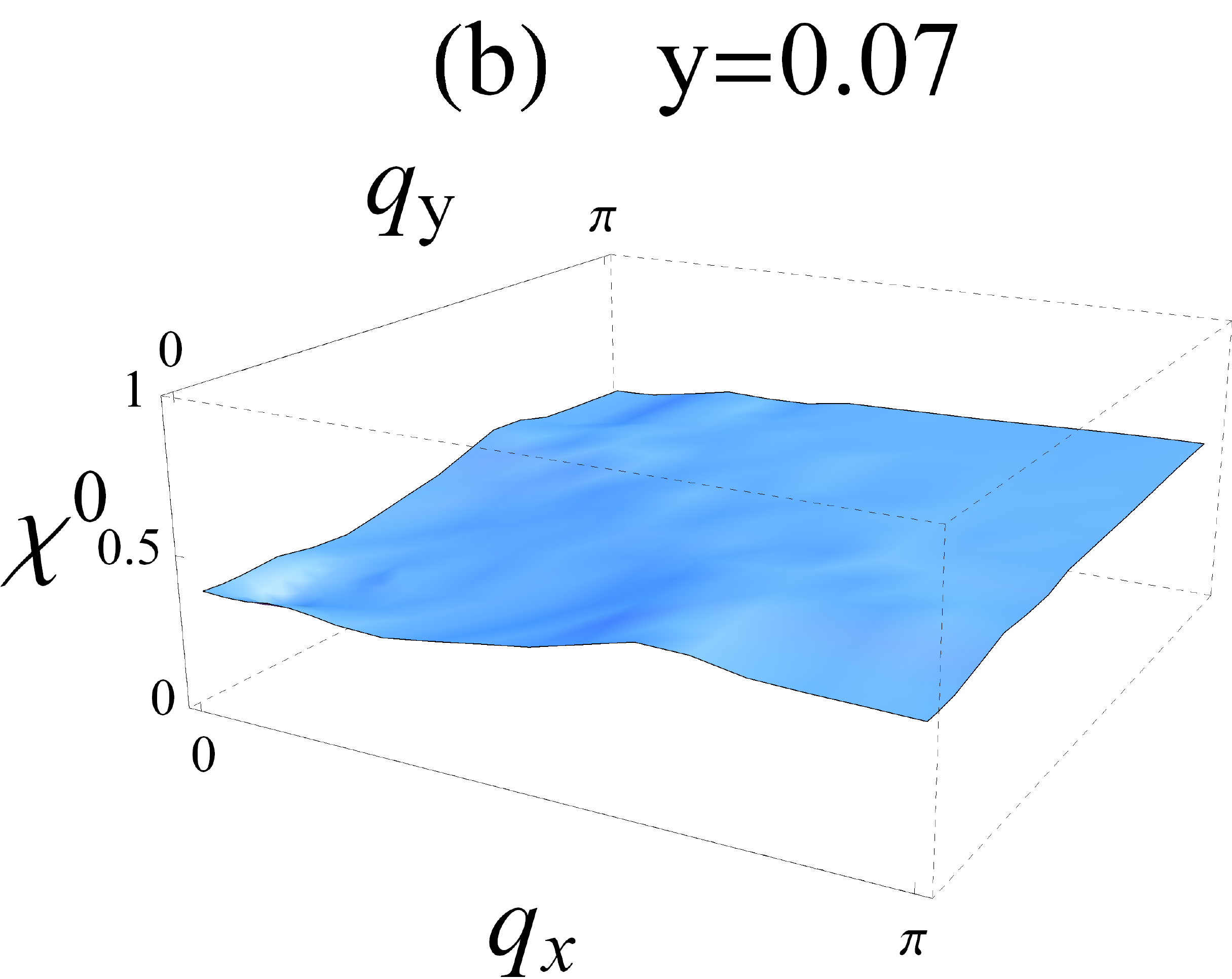}
\includegraphics[width=0.65\columnwidth]{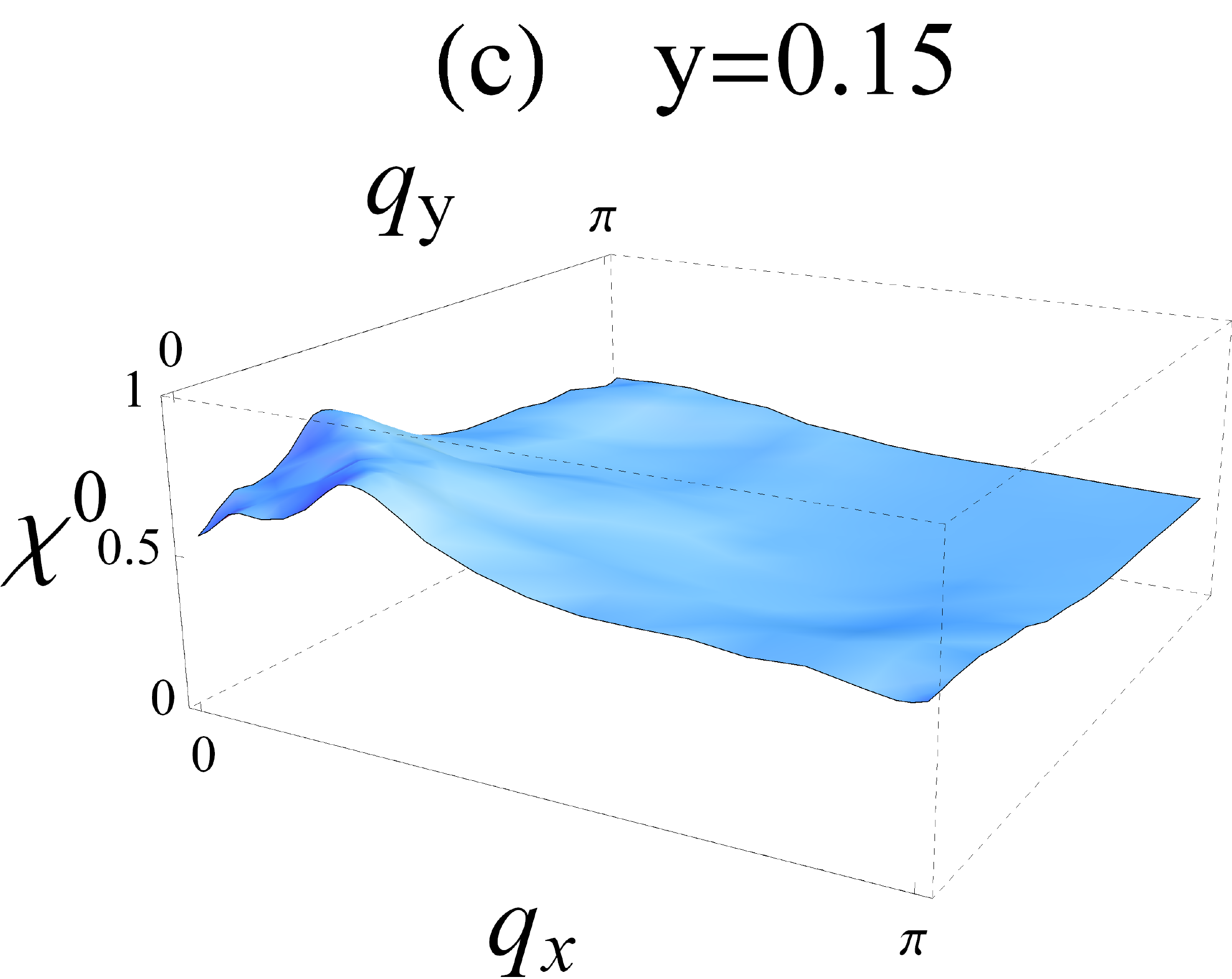} \\[0.5cm] \includegraphics[width=0.65\columnwidth]{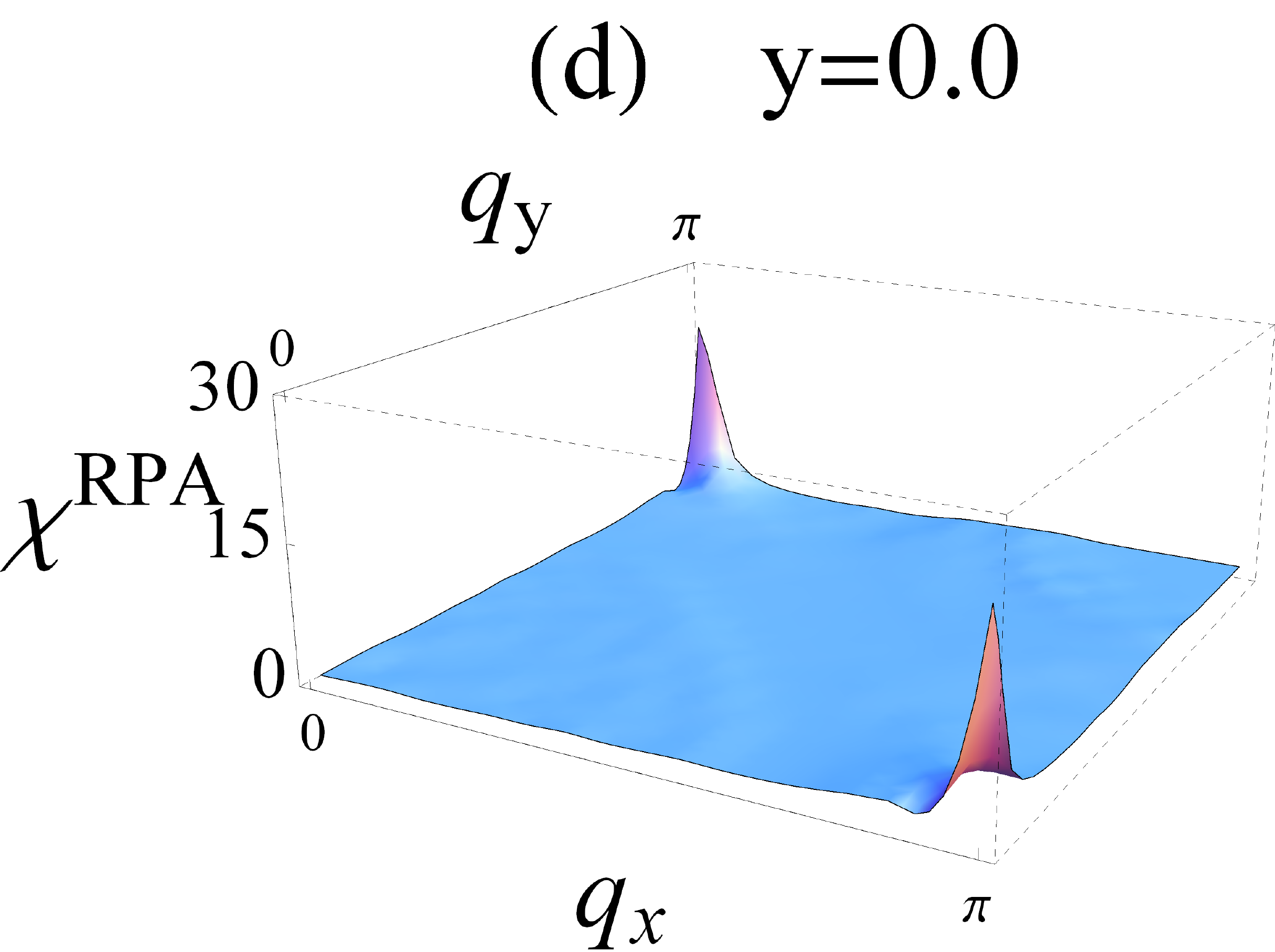}
\includegraphics[width=0.65\columnwidth]{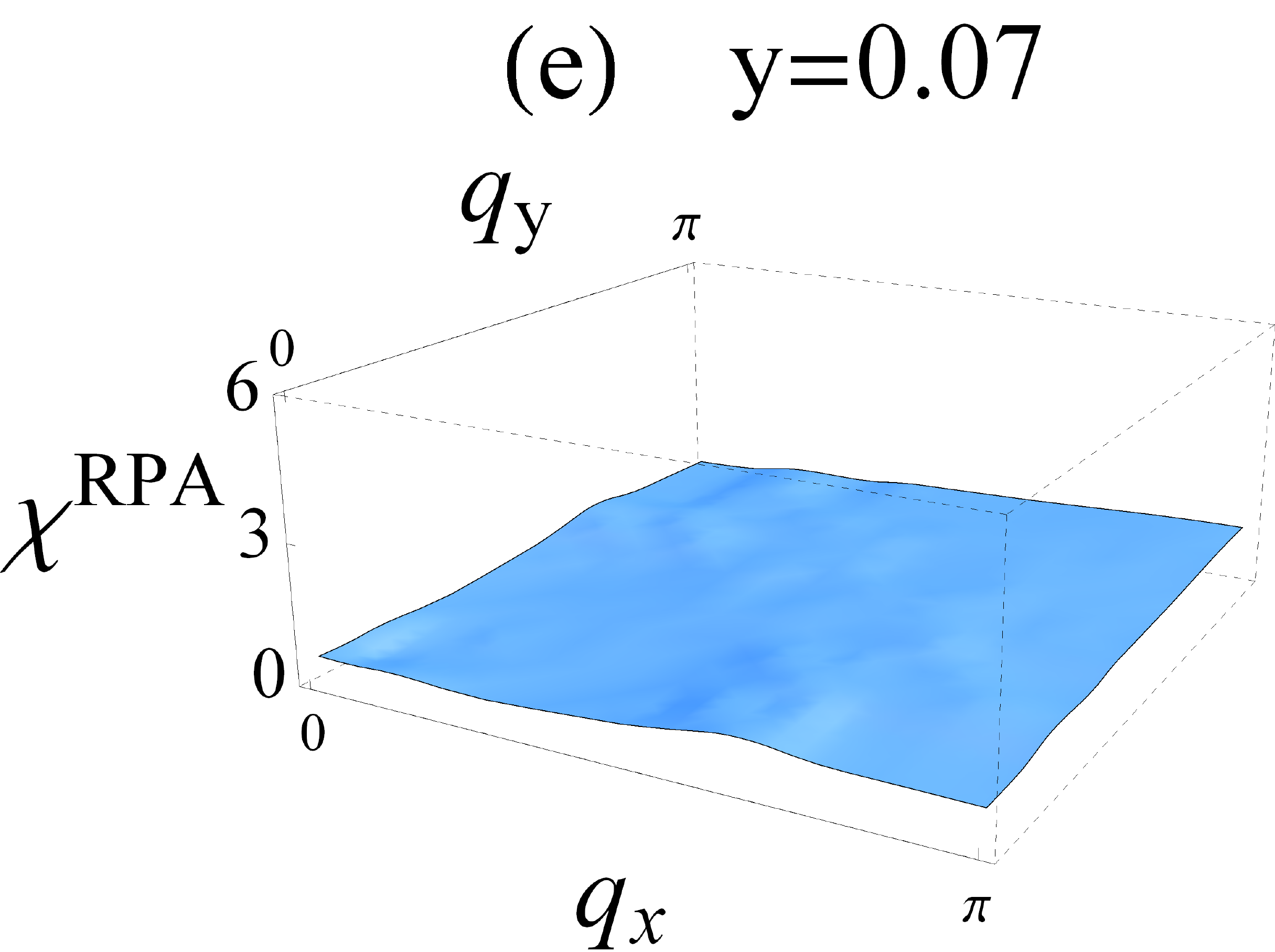} \includegraphics[width=0.65\columnwidth]{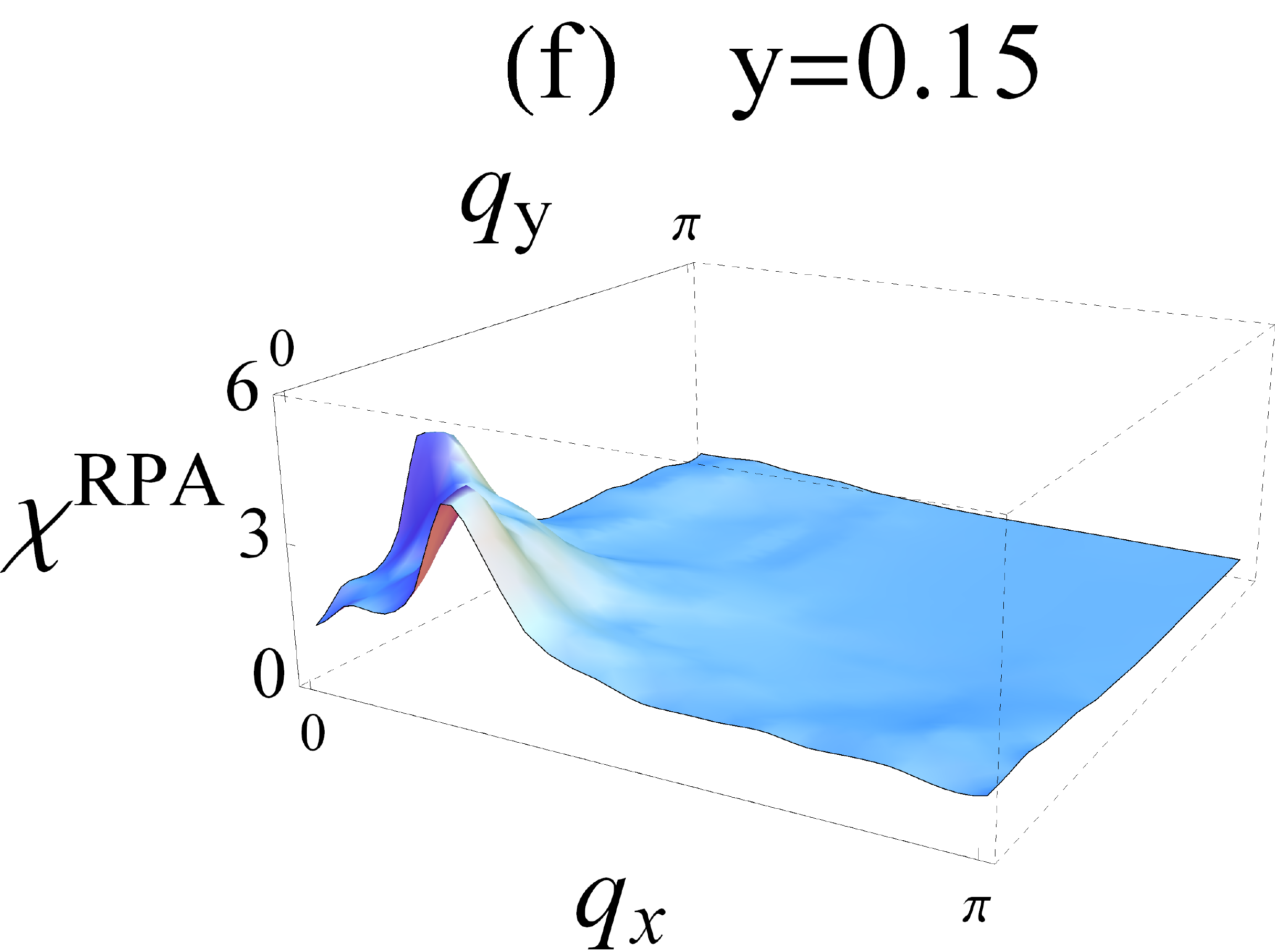}
\caption{(Color online) (a)-(c) The bare static spin susceptibility $\chi^{0}(\mathbf{q},0)$
and (d)-(f) the RPA spin susceptibility $\chi^{\mathrm{RPA}}(\mathbf{q},0)$
calculated using the TBM after projecting
out the $x^{2}-y^{2}$ and $3z^{2}-r^{2}$ orbitals. The concentration
of excess Fe is (a) and (d) $y=0.0$; (b) and (e) $y=0.07$; (c) and
(f) $y=0.15$. The RPA spin susceptibility is calculated for $U=1.0$
eV and $J_{H}=U/5$.}
\end{figure*}

\subsection{Bare susceptibility}
In the multi-orbital model under consideration, the matrix elements
of the bare spin susceptibility can be calculated from the corresponding
Matsubara spin-spin correlation function of conduction electrons:
\begin{eqnarray}
 &  & \chi_{aa'bb'}^{0}({\mathbf{q}},\imath\nu_{n})\label{sus1}\\
 &  & =-\frac{T}{N}\sum_{{\mathbf{k}},\imath\omega_{n}}G_{ab}({\mathbf{k}},\imath\omega_{n})G_{a'b'}({\mathbf{k}}+{\mathbf{q}},\imath\omega_{n}+\imath\nu_{n})\nonumber
\end{eqnarray}
 where $a,a',b,b'$ are orbital indices, and the spectral representation
of the multi-orbital Green's function is given by
\begin{eqnarray}
G_{ab}({\mathbf{k}},\imath\omega_{n})=\sum_{\nu}\frac{u_{\nu}^{a}({\mathbf{k}})(u_{\nu}^{b}({\mathbf{k}}))^{*}}{\imath\omega_{n}-E_{\nu}({\mathbf{k}})}~.\label{sus2}
\end{eqnarray}
Here, the
 matrix elements $u_{\nu}^{a}({\mathbf{k}})$ are the components of
the eigenvectors resulting from the diagonalization of the five-orbital
TBM and $E_{\nu}({\mathbf{k}})$
are the eigenvalues describing the resulting bands characterized by
the band index $\nu$. The retarded bare susceptibility is then obtained
by summing over the Matsubara frequency and setting $\imath\nu_{n}\rightarrow\omega+\imath\delta$:
\begin{eqnarray}
\chi_{aa'bb'}^{0}(\mathbf{q},0) & = & -\frac{1}{N}\sum_{\mathbf{k},\nu,\nu'}\frac{u_{\nu\mathbf{k}}^{a}(u_{\nu\mathbf{k}}^{b})^{*}u_{\nu'(\mathbf{k}+\mathbf{q})}^{b'}(u_{\nu'(\mathbf{k}+\mathbf{q})}^{a'})^{*}}{E_{\nu'}(\mathbf{k}+\mathbf{q})-E_{\nu}(\mathbf{k})}\nonumber \\
 &  & \times\left(f(E_{\nu'}(\mathbf{k}+\mathbf{q}))-f(E_{\nu}(\mathbf{k}))\right)~,
\end{eqnarray}
 where
\[
f(E_{\nu}(\mathbf{k}))=\frac{1}{e^{\frac{E_{\nu}(\mathbf{k})-\mu}{T}}+1}
\]
 denotes the Fermi distribution function.

In Fig. 3 (a)-(c) we show the results for the total bare spin susceptibility
$\chi^{0}(\mathbf{q},0)=\frac{1}{2}\chi_{aabb}^{0}(\mathbf{q},0)$
where
 the $x^{2}-y^{2}$ and $3z^{2}-r^{2}$ orbitals have been projected
out by setting all components of the eigenvectors $u_{\nu}^{x^{2}-y^{2}}({\mathbf{k}})$
and $u_{\nu}^{3z^{2}-r^{2}}({\mathbf{k}})$ equal to zero. We note
that the results do not change significantly if these orbitals are
included, since they do not contribute to the Fermi surface.

The bare susceptibility $\chi^{0}(\mathbf{q})$ is overall rather
flat with small peaks at $(0,\pi)$ and $(\pi,0)$ for $y=0$ (Fig.
3 (a)), almost featureless for $y=0.07$ (Fig. 3 (b)), and with a
wide region of enhanced fluctuations in the vicinity of the $\Gamma$
point for $y=0.15$ (Fig. 3 (c)). All these features are displayed
in Fig. 4 (a), where we plot the bare spin susceptibilities for different
values of $y$ along the high-symmetry path $\Gamma=(0,0)\rightarrow X=(\pi,0)\rightarrow M=(\pi,\pi)\rightarrow\Gamma=(0,0)$.
Red, green and blue lines correspond to $y=0.0$, 0.07 and 0.15,
respectively. As in Figs. 3 (a)-(c), we see that for $y=0.0$ the bare
susceptibility shows enhanced fluctuations peaked at the $X=(\pi,0)$
point, whereas for $y=0.15$ the magnetic spectral weight is shifted
to the vicinity of $\Gamma=(0,0)$ point. For intermediate values
of $y$, the bare susceptibility is basically featureless.

\subsection{RPA susceptibility}

The effect of correlations among the itinerant electrons on the spin
susceptibility can be taken into account in the framework of the RPA.
The RPA spin susceptibility can be obtained using the Dyson equation:\cite{graser09}
\begin{eqnarray}
\chi_{aa'bb'}^{{\rm RPA}}(\mathbf{q},\omega) & = & \chi_{aa'bb'}^{0}(\mathbf{q},\omega)+\\
 &  & \chi_{aa'cc'}^{0}(\mathbf{q},\omega)\, V_{cc'dd'}\,\chi_{dd'bb'}(\mathbf{q},\omega)~,\nonumber
\end{eqnarray}
 where the matrix elements $V_{cc'dd'}$ take into account all electron
correlations in the RPA. The interaction parameters which give the
strongest contributions are those from the matrix elements $V_{cccc}=U$,
$V_{ccdd}=J_{H}$, $V_{cddc}=J_{H}$, and $V_{cdcd}=U-2J_{H}$. All
other matrix elements are set to 0. In our calculations we have set
$U=1.0$ eV and $J_{H}=U/5$. Both the Coulomb repulsion and the Hund's
coupling are well inside the range of the interaction parameters previously
considered in the literature: the lowest estimate of $J_{H}/U=0.15$
was used in Ref.\cite{lanata13} and the upper limit of $J_{H}/U=0.25$
was considered in several works, e.g., in Refs.\cite{brydon11,ding13}.
Selecting $J_{H}$ inside this range ensures that an electron added
to an undoped site pays more energy to Coulomb repulsion than it wins
from the Hund's rule, i.e., that the onsite interaction energy suppresses
charge fluctuations rather than enhancing them.\cite{brydon11}

In Figs. 3 (d)-(f), we present the static RPA spin susceptibility for
different values of $y$. Overall, $\chi^{{\rm RPA}}(\mathbf{q},0)$
shows a significant enhancement due to interactions -- notice, for
instance, the different scales used in Figs. 3 (a)-(f).
The RPA susceptibility for $y=0.0$ is shown in Fig. 3 (d). The small
peaks at the wave vectors $(\pi,0)$ and $(0,\pi)$ observed in the
bare susceptibility now display a nearly-diverging behavior. However,
similarly to the bare susceptibility, $\chi^{{\rm RPA}}(\mathbf{q},0)$
for $y=0.0$ does not display significant spin fluctuations near the
($\pm\pi/2,\pm\pi/2$) points, corresponding to the ordering vectors
of the experimentally observed magnetic order.

With increasing $y$ (see Fig. 3 (e) and Fig. 3 (f)), we observe significant
changes in the overall structure of the spin susceptibility. As it
is particularly seen in Fig. 4 (b), where we show the RPA spin susceptibility
along the main symmetry directions, the susceptibility near $(\pi,0)$
and $(0,\pi)$ rapidly decreases with increasing $y$, and the dominant
magnetic response shifts to the vicinities of the $\Gamma$ point.
This is consistent with the loss of nesting features in the Fermi
surface. For $y=0.07$ (Fig. 3 (e)), the susceptibility is almost
uniform across the whole Brillouin zone. For higher values of $y$
(see Fig. 3 (f)), the susceptibility shows dominant but not diverging
behavior in the central part of Brillouin zone close to the $\Gamma$
point.

\begin{figure}
\label{fig4} \includegraphics[width=0.9\columnwidth]{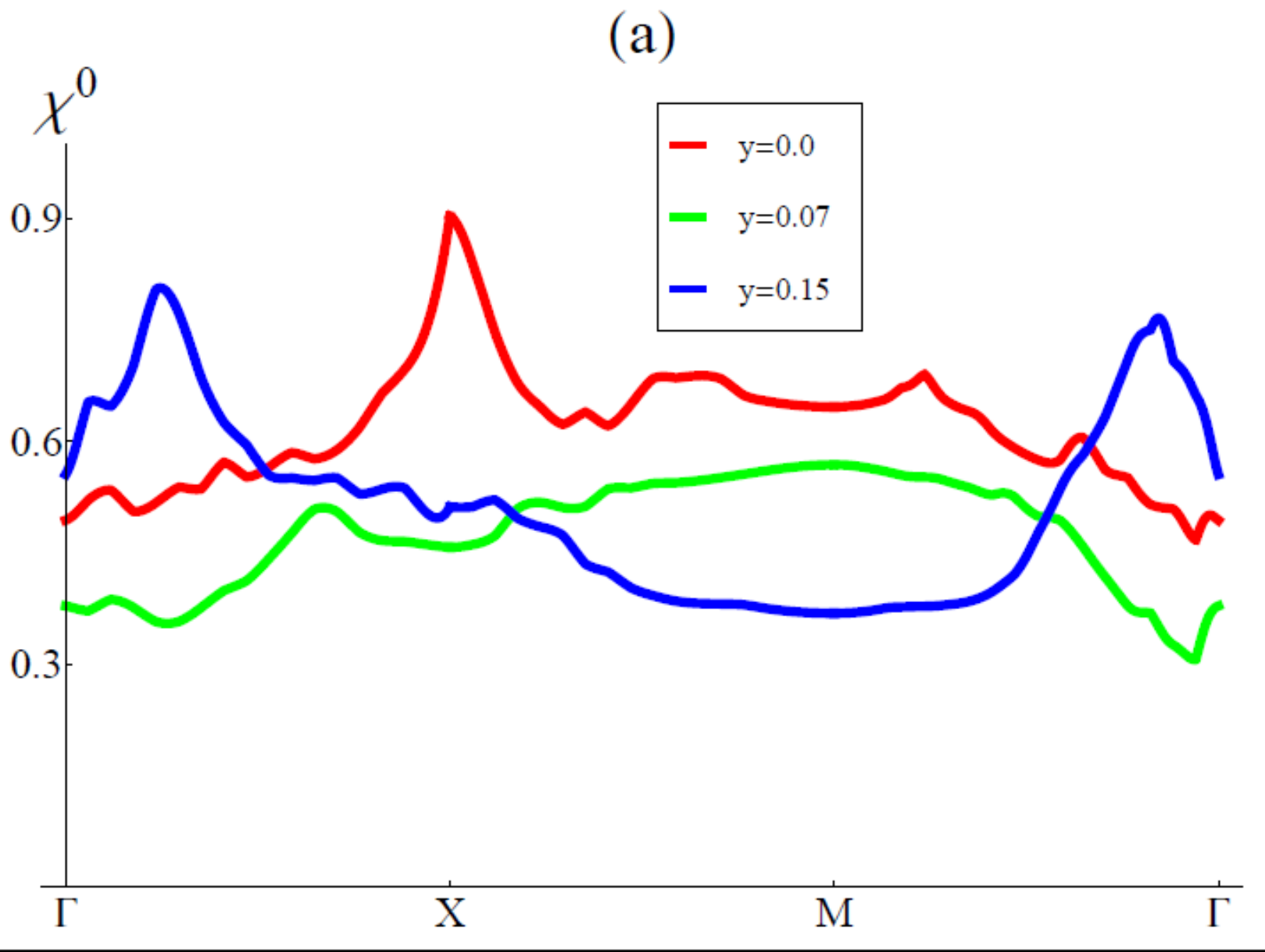} \includegraphics[width=0.9\columnwidth]{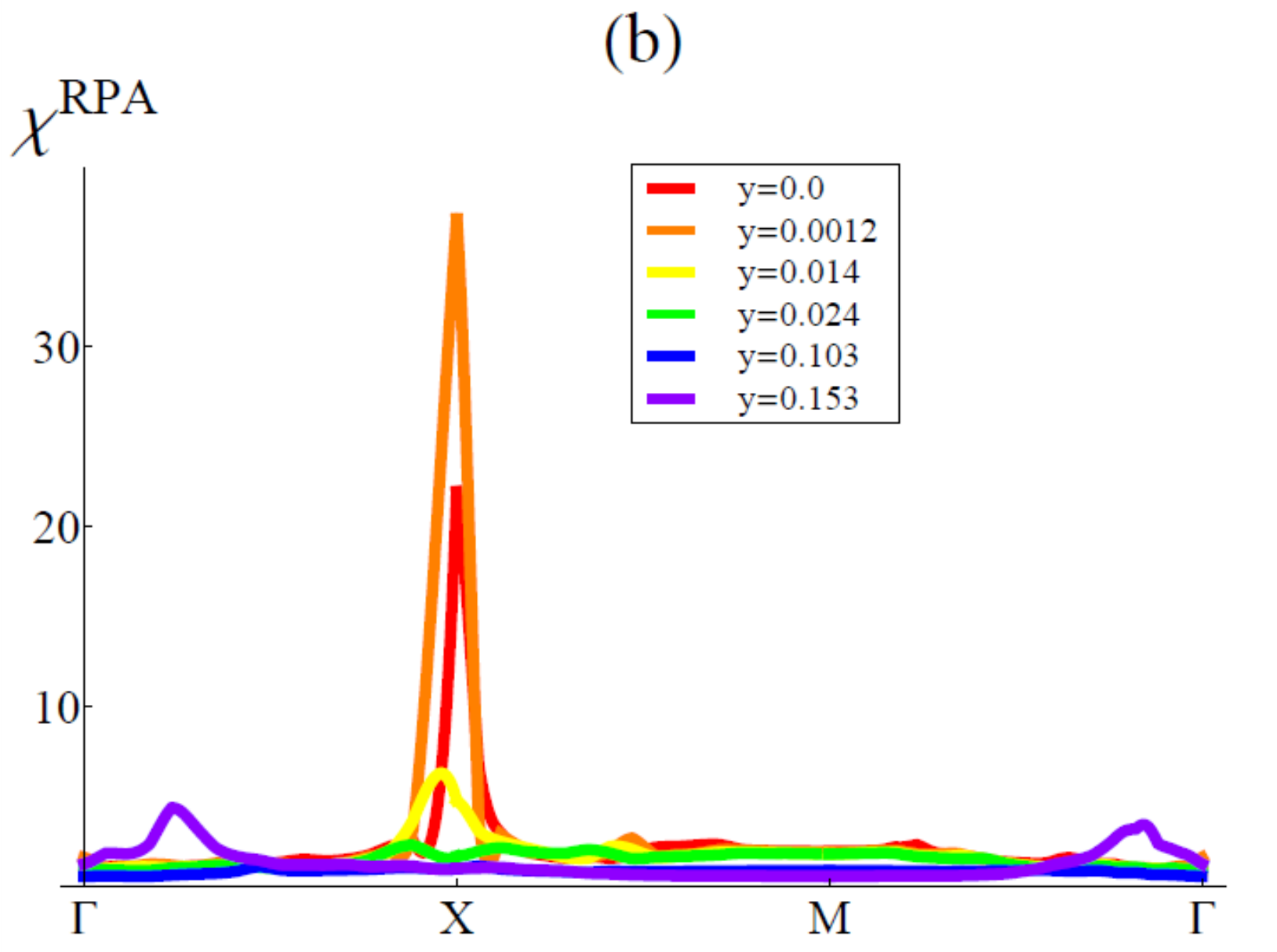}
\caption{(Color online) (a) Cut of the bare spin susceptibility along a high-symmetry
path. Red, green and blue lines correspond to $y=0.0$, 0.07
and 0.15, respectively. (b) Cut of the RPA enhanced spin susceptibility
along the same high-symmetry path. The legend provides the $y$ values
of each line. We use the following notations: $\Gamma=(0,0)$, $X=(\pi,0)$,
and $M=(\pi,\pi)$. The RPA spin susceptibility is calculated for
$U=1.0$ eV and $J_{H}=U/5$.}
\end{figure}

\section{The RKKY exchange integrals}

\label{RKKY_sec}

Equation (\ref{RKKYcouling}) shows that the RKKY exchange integrals $J_{ij}^{\mathrm{RKKY}}$
are proportional to the static magnetic susceptibility of the multi-orbital
conduction electrons. Thus, the changes in the spin susceptibility
promoted by the increase in the
 concentration of interstitial Fe discussed in the previous section
will lead to changes in $J_{ij}^{\mathrm{RKKY}}$ and, according to
Eq. (\ref{exchange-model}), to changes in the effective Hamiltonian
of the localized spins.

\begin{figure*}
\label{fig5} \includegraphics[width=1\columnwidth]{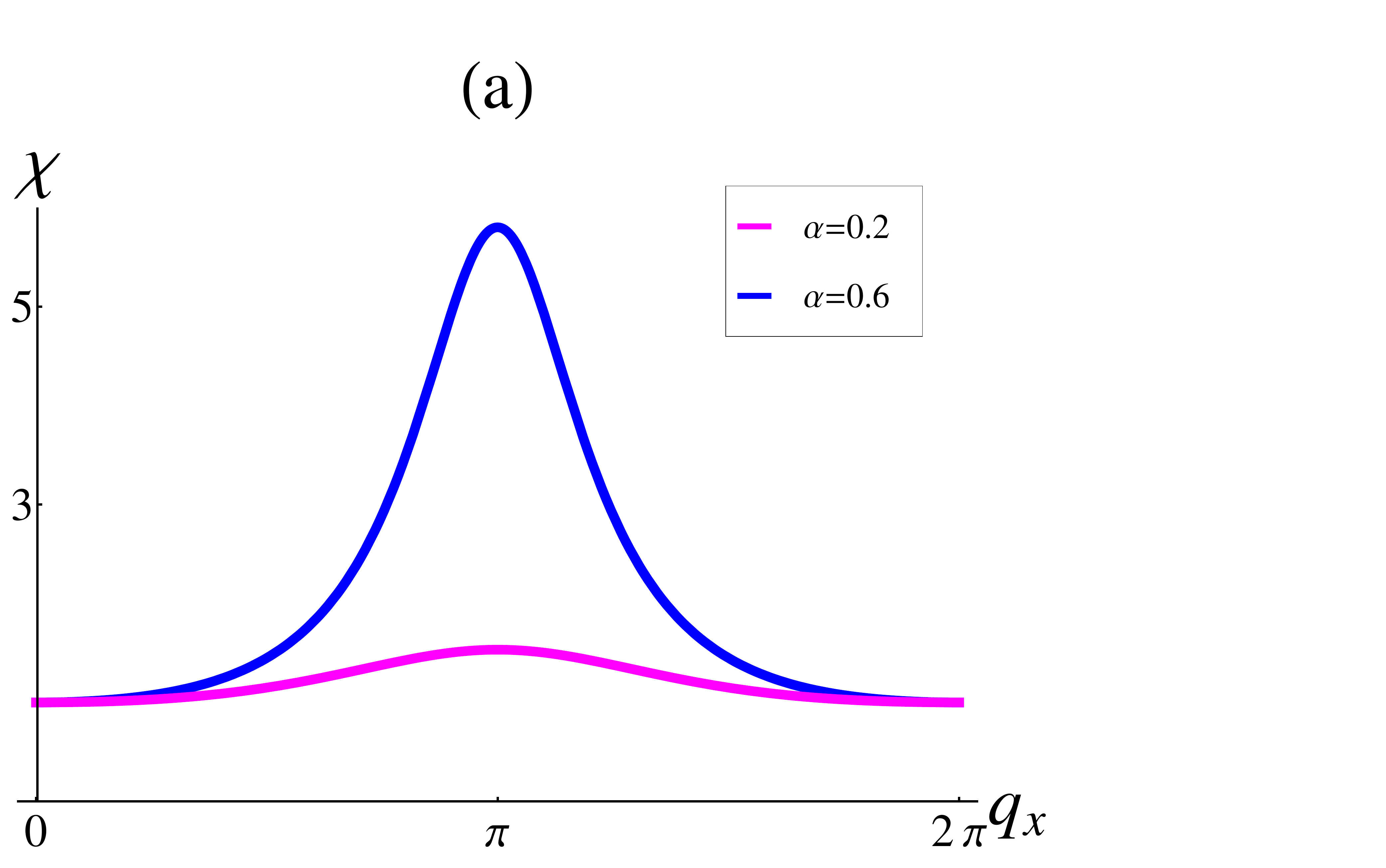} \includegraphics[width=1\columnwidth]{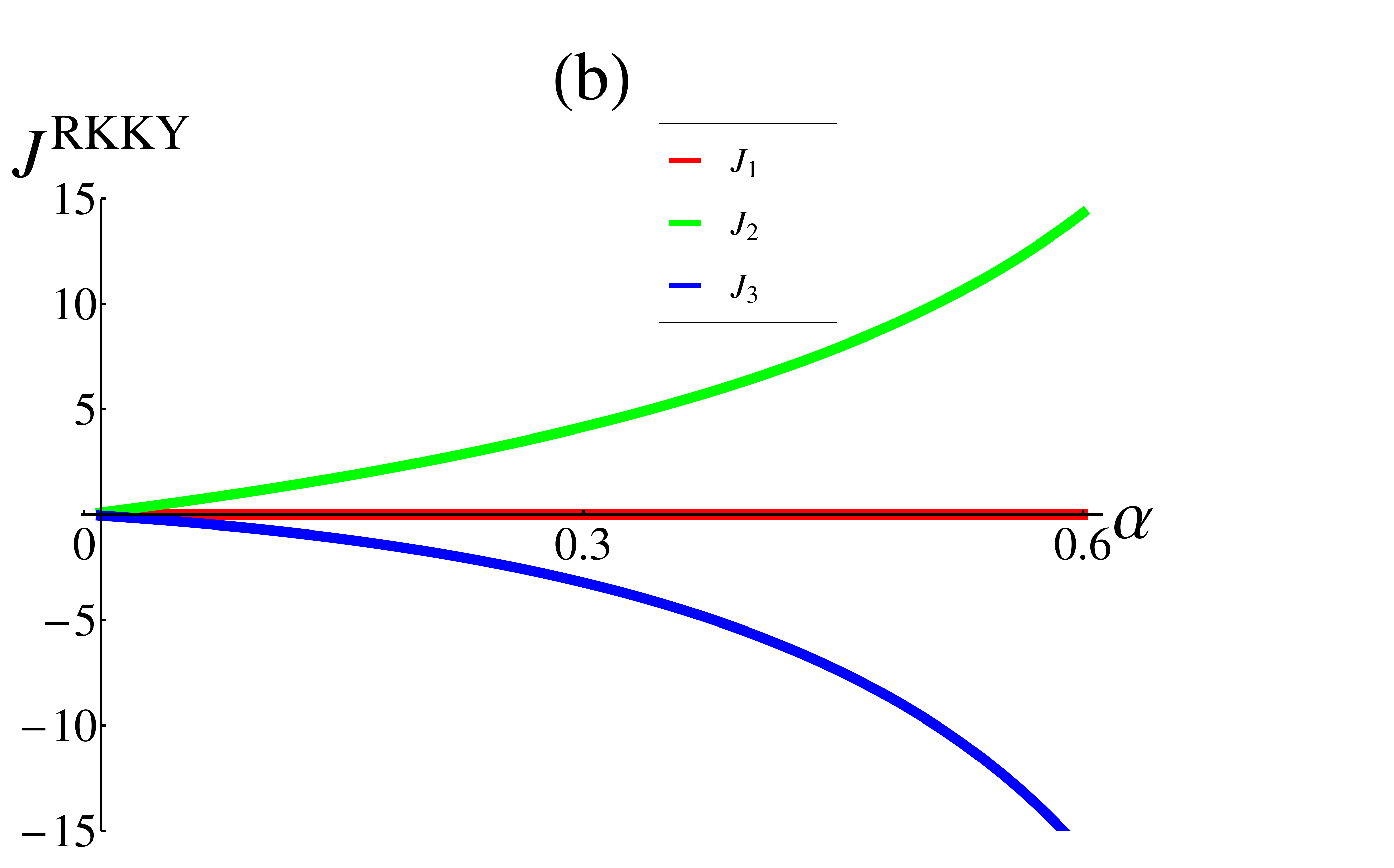}
\caption{(a) Static magnetic susceptibility $\chi\left(q_{x},0\right)$ of
the toy model (\ref{mag_suscep}) for $\alpha=0.2$ (magenta) and
$\alpha=0.6$ (blue). (b) $J^{\mathrm{RKKY}}$ (in units of $J_{H}^{2}\chi_{0}$)
of the same toy model as function of the $\left(\pi,0\right)/\left(0,\pi\right)$
peak intensity $\alpha$. $J_{1}^{\mathrm{RKKY}}$, $J_{2}^{\mathrm{RKKY}}$
and $J_{3}^{\mathrm{RKKY}}$ are shown by red, green, and blue lines,
respectively.}
\end{figure*}

\subsection{Toy model for the RKKY interaction}

As shown in Fig. 4 (b), one of the main effects of increasing $y$
on the spin susceptibility is to suppress the nesting-induced peaks
at momenta $\left(\pi,0\right)$ and $\left(0,\pi\right)$. To shed
light on how these changes are translated in changes of the RKKY exchange
interactions, we first consider a simple toy model in which the spin
susceptibility is given by the phenomenological expression \cite{Paul11_PRL}:
\begin{eqnarray}
\chi^{-1}\left(\mathbf{q}\right)=\frac{1+\alpha\left[\cos q_{x}\cos q_{y}-\frac{1}{8}\left(\cos2q_{x}+\cos2q_{y}\right)\right]}{\chi_{0}\left(1+\frac{3}{4}\alpha\right)}\label{mag_suscep}
\end{eqnarray}

Here, $\chi_{0}$ sets the overall scale for magnetic fluctuations
and $\alpha<1$ is a parameter that controls the height and width
of the peaks at $\left(\pi,0\right)/\left(0,\pi\right)$. The overall
amplitude of the magnetic susceptibility is kept unchanged by the
term $\left(1+\frac{3}{4}\alpha\right)$ in the denominator. In Fig. 5
(a) we plot this phenomenological static magnetic susceptibility along
the $X-\Gamma$ direction, $\chi\left(q_{x},0\right)$ for $\alpha=0.2$
(magenta) and $\alpha=0.6$ (blue), illustrating how the peak at $\left(\pi,0\right)$
decreases with decreasing $\alpha$. Thus, decreasing $\alpha$ mimics
the effect of increasing $y$ in Fig. 4 (b) (at least for small $y$).

Next we compute the RKKY interactions as a function of $\alpha$.
Taking the Fourier transforms, we obtain:
\begin{eqnarray}
J_{1}^{\mathrm{RKKY}} & = & -J_{H}^{2}\sum_{\mathbf{q}}\chi\left(\mathbf{q}\right)\left(\cos q_{x}+\cos q_{y}\right)\\
J_{2}^{\mathrm{RKKY}} & = & -J_{H}^{2}\sum_{\mathbf{q}}\chi\left(\mathbf{q}\right)\cos q_{x}\cos q_{y}\\
J_{3}^{\mathrm{RKKY}} & = & -J_{H}^{2}\sum_{\mathbf{q}}\chi\left(\mathbf{q}\right)\left(\cos2q_{x}+\cos2q_{y}\right)\label{RKKY_toy}
\end{eqnarray}

A straightforward evaluation gives the results shown in Fig. 5(b).
While $J_{1}^{\mathrm{RKKY}}=0$ for all values of $\alpha$, we note
that $J_{2}^{\mathrm{RKKY}}>0$ and $J_{3}^{\mathrm{RKKY}}<0$, with
$J_{2}^{\mathrm{RKKY}}\approx\left|J_{3}^{\mathrm{RKKY}}\right|$.
As the peak intensity decreases (i.e. as $\alpha$ decreases), the
absolute values of $J_{2}^{\mathrm{RKKY}}$ and $J_{3}^{\mathrm{RKKY}}$
decrease strongly.

This behavior can be understood in a straightforward way by noting
that only the structure factors of $J_{2}^{\mathrm{RKKY}}$ and $J_{3}^{\mathrm{RKKY}}$
in Eq. (\ref{RKKY_toy}) match the Fourier components of the magnetic
susceptibility in Eq. (\ref{mag_suscep}). To make this argument even
more transparent, consider an even simpler model for the magnetic
susceptibility consisting of a constant background plus peaks at $\mathbf{Q}_{X}=\left(\pi,0\right)$
and $\mathbf{Q}_{Y}=\left(0,\pi\right)$:
\begin{eqnarray}
\frac{\chi\left(\mathbf{q}\right)}{\chi_{0}}=1+\alpha\left[\delta\left(\mathbf{q}-\mathbf{Q}_{X}\right)+\delta\left(\mathbf{q}-\mathbf{Q}_{Y}\right)\right]\label{simpler}
\end{eqnarray}

It is straightforward to obtain:
\begin{eqnarray*}
J_{1}^{\mathrm{RKKY}}/(J_{H}^{2}\chi_{0}) & = & -2\alpha\left(\cos\pi+\cos0\right)=0\\
J_{2}^{\mathrm{RKKY}}/(J_{H}^{2}\chi_{0}) & = & -2\alpha\left(\cos\pi\cos0\right)=4\alpha\\
J_{3}^{\mathrm{RKKY}}/(J_{H}^{2}\chi_{0}) & = & -2\alpha\left(\cos2\pi+\cos0\right)=-4\alpha
\end{eqnarray*}
 in agreement with Fig. 5 (b). Thus, this toy model shows that peaks
at $\left(\pi,0\right)$ and $\left(0,\pi\right)$ in the itinerant
susceptibility induce local-spin interactions only between second
and third neighbors, without affecting the
first-neighbors interaction. The latter is, however, very sensitive
to the fluctuations peaked at different wave vectors, as we will see
in more realistic calculations in the next subsection, in particular
to the fluctuations with small-$\mathbf{q}$ vectors.

\subsection{RKKY interactions computed within the TBM}

\begin{figure}
\label{fig6} \includegraphics[width=0.95\columnwidth]{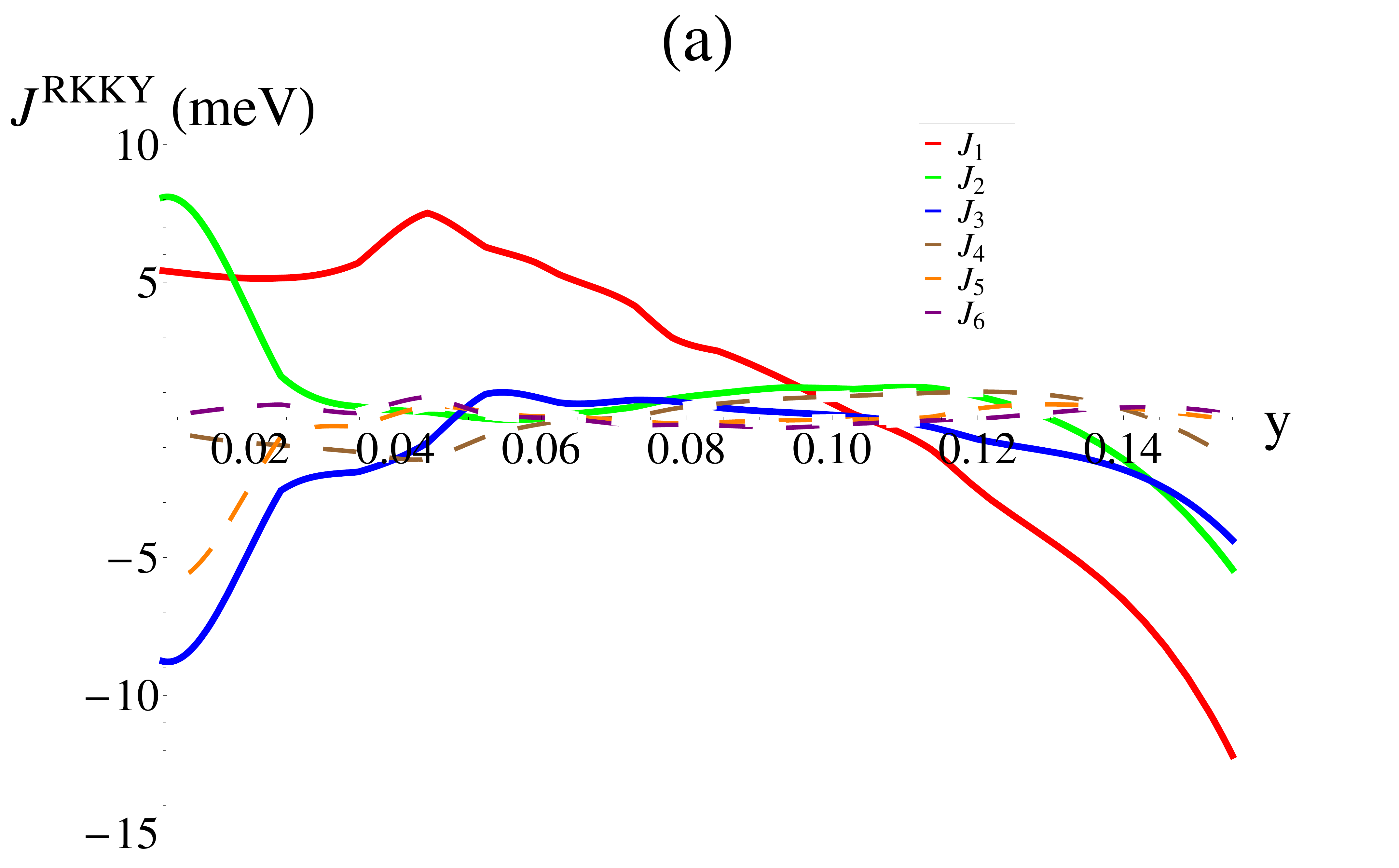} \includegraphics[width=0.95\columnwidth]{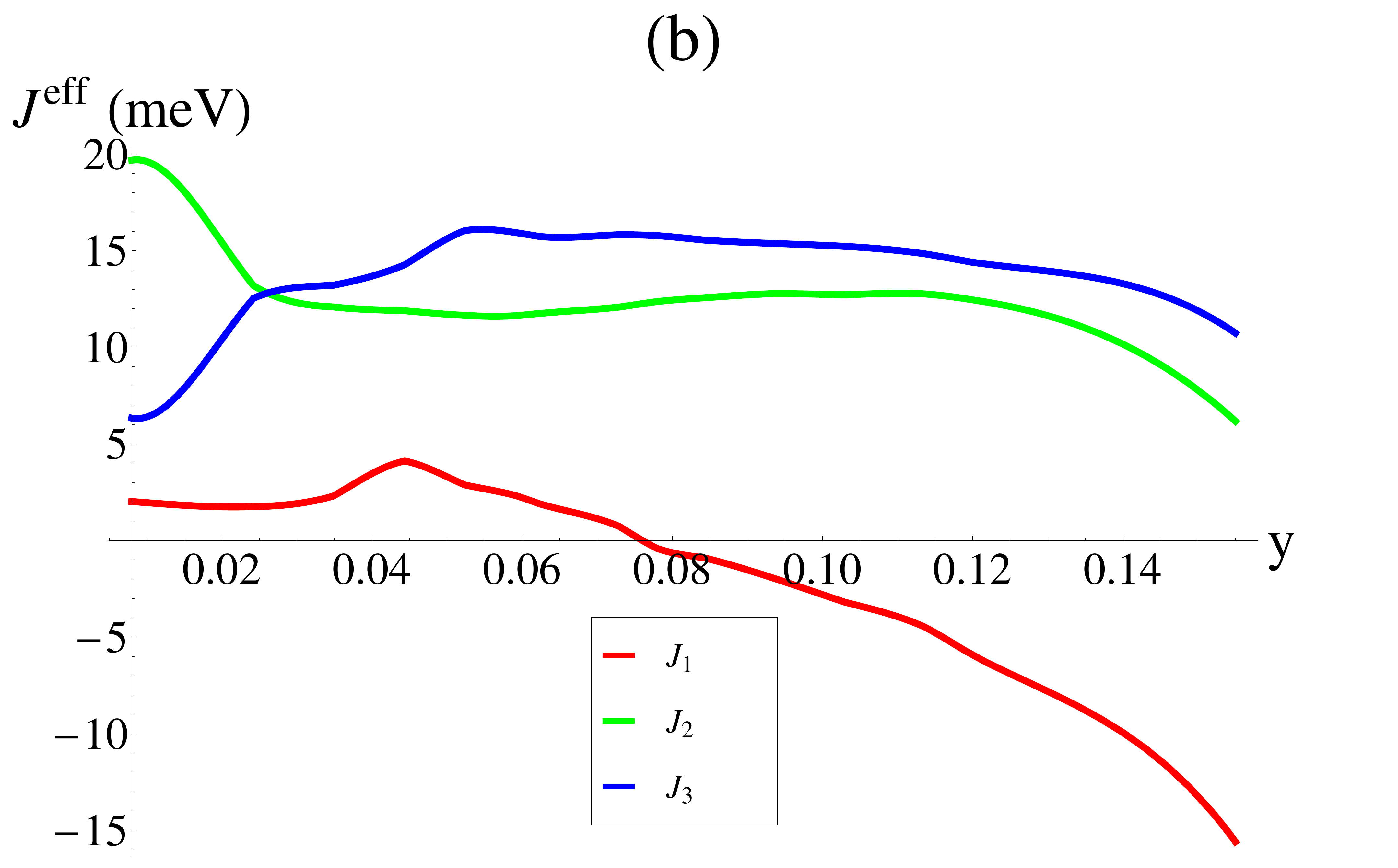}
\caption{ (Color online) (a) The evolution of $J_{1}^{\mathrm{RKKY}}$ (red solid line), $J_{2}^{\mathrm{RKKY}}$
(green solid line), $J_{3}^{\mathrm{RKKY}}$ (blue solid line) with
increasing concentration of excess of interstitial Fe atoms $y$.
The other neighbor interactions are $J_{4}^{\mathrm{RKKY}}$ (brown
dashed line), $J_{5}^{\mathrm{RKKY}}$ (orange dashed line), $J_{6}^{\mathrm{RKKY}}$
(purple dashed line). (b) The evolution of the effective exchange
couplings $J_{ij}^{\mathrm{eff}}$, where $J_{1}^{\mathrm{eff}}=-3.4+J_{1}^{\mathrm{RKKY}}(y)$
(red line), $J_{2}^{\mathrm{eff}}=11.6+J_{2}^{\mathrm{RKKY}}(y)$
(green line), and $J_{3}^{\mathrm{eff}}=15.1+J_{3}^{\mathrm{RKKY}}(y)$
(blue line). All interactions are given in meV/$S^2$.}
\end{figure}

We now compute the RKKY interactions numerically using the results
for the susceptibility obtained in Sec. IV. For completeness, we compute
the RKKY interactions up to the sixth-nearest neighbors for different
concentrations of excess interstitial Fe, $y$. In Fig. 6 (a) we plot
$J_{1}^{\mathrm{RKKY}}$ , $J_{2}^{\mathrm{RKKY}}$ and $J_{3}^{\mathrm{RKKY}}$
-- shown, correspondingly, in red, green and blue solid lines -- and
$J_{4}^{\mathrm{RKKY}}$, $J_{5}^{\mathrm{RKKY}}$ and $J_{6}^{\mathrm{RKKY}}$
-- shown, correspondingly, in brown, orange and purple dashed lines.
$J_{4}^{\mathrm{RKKY}}-J_{6}^{\mathrm{RKKY}}$ interactions are  small almost for almost all values of $y$, except
$J_{5}^{\mathrm{RKKY}}$  at interstitial concentrations  $y<0.015$.  However, as Fe$_{1+y}$Te crystals have not been yet grown with $y<0.015$,  we  will not discuss the possible effects of this term.

For small but physical values of $y$, the largest interactions are $J_{2}^{\mathrm{RKKY}}$
and $J_{3}^{\mathrm{RKKY}}$, whereas for intermediate and larger
values of $y$, $J_{1}^{\mathrm{RKKY}}$ dominates. Because of this,
and since in our bare local model (\ref{hamloc}) only $J_{1}$, $J_{2}$,
and $J_{3}$ have non-zero values, hereafter we neglect the RKKY contributions
beyond third-nearest neighbors.

Much of the behavior of the RKKY interactions for small values of
$y$ can be understood within the toy model discussed in the previous
section. As expected, we obtain large antiferromagnetic $J_{2}^{\mathrm{RKKY}}$
and ferromagnetic $J_{3}^{\mathrm{RKKY}}$ interactions in the regime
where the susceptibility $\chi^{{\rm RPA}}(\mathbf{q},0)$ has strong
peaks near the $(\pi,0)$ and $(0,\pi)$ points ($y\lesssim0.025$,
see Fig. 4 (b)). As these peaks are suppressed, the absolute values
of both interaction parameters are reduced, in agreement with decreasing
$\alpha$ in Fig. 5 (b). At the same time, the small antiferromagnetic
$J_{1}^{\mathrm{RKKY}}$ interaction remains nearly unchanged, reflecting
the fact that it is unaffected by the fluctuations near $(\pi,0)$
and $(0,\pi)$. In contrast, fluctuations in broader regions of the
Brillouin zone,
such as in the vicinities of the $M$ and the $\Gamma$ points, yield
the leading contributions to the $J_{1}^{\mathrm{RKKY}}$ interaction.

By increasing the concentration of interstitial Fe atoms beyond $y\approx0.025$,
the $J_{1}^{\mathrm{RKKY}}$ interaction becomes the dominant one,
as the large peaks of the itinerant susceptibility $\chi^{{\rm RPA}}(\mathbf{q},0)$
near $(\pi,0)$ and $(0,\pi)$ quickly disappear, rendering $J_{2}^{\mathrm{RKKY}}$
and $J_{3}^{\mathrm{RKKY}}$ small. However, for $y\gtrsim0.05$,
the antiferromagnetic $J_{1}^{\mathrm{RKKY}}$ interaction starts
being suppressed, and eventually changes sign and becomes ferromagnetic
for $y\gtrsim0.1$. Comparison to the behavior of the RPA susceptibility
in Fig. 4 (b) reveals that this change can be attributed to the reduction
of the broad fluctuations present around the $M$ point, followed
by the subsequent enhancement of fluctuations near the $\Gamma$ point.
Indeed, according to the form factor of $J_{1}^{\mathrm{RKKY}}$ in
Eq. (\ref{RKKY_toy}), fluctuations around $M=\left(\pi,\pi\right)$
yield an antiferromagnetic $J_{1}^{\mathrm{RKKY}}$, whereas fluctuations
around $\Gamma=\left(0,0\right)$ yield a ferromagnetic $J_{1}^{\mathrm{RKKY}}$.

\subsection{Derivation of the effective couplings}

Having calculated $J_{ij}^{\mathrm{RKKY}}$ as function of the Fe
excess concentration, it is now straightforward to compute the $y$-dependent
effective couplings $J_{ij}^{\mathrm{eff}}$ between the localized
moments in Eq. (\ref{exchange-model}), whose solution determines
the magnetic ground state. To this end, we first need to establish
the bare Heisenberg exchange couplings $J_{ij}$, which do not change
with increasing Fe excess. The values of the effective couplings $J_{ij}^{\mathrm{eff}}$
at $y\simeq0$ were computed via first-principles electronic structure
calculations by Ma \textit{et al}.\cite{Ma09} Up to the third neighbors,
these values normalized on the value of spin $S$ are equal to $J_{1}^{\mathrm{eff}}=$2.1 meV/$S^2$, $J_{2}^{\mathrm{eff}}=$15.8
meV/$S^2$, $J_{3}^{\mathrm{eff}}=$10.1 meV/$S^2$. Because even for $y\simeq0$,
Fe$_{1+y}$Te has both itinerant and localized electrons, these effective
couplings contain both the $J_{ij}^{\mathrm{RKKY}}$ interactions
and the bare Heisenberg exchange couplings $J_{ij}$. In order to
decompose these contributions, we simply subtract the RKKY interactions
computed by us at $y=0$ from the values of the super-exchange interactions
derived by Ma \textit{et al}.~\cite{Ma09} This procedure yields
the bare Heisenberg exchange couplings $J_{1}=-3.4$ meV/$S^2$, $J_{2}=11.6$
meV/$S^2$, and $J_{3}=15.1$ meV/$S^2$.

In Fig. 6 (b), we plot $J_{ij}^{\mathrm{eff}}(y)=J_{ij}+J_{ij}^{\mathrm{RKKY}}(y)$
as a function of $y$. We note that the effective couplings $J_{2}^{\mathrm{eff}}$
and $J_{3}^{\mathrm{eff}}$ remain antiferromagnetic for all $y$,
while $J_{1}^{\mathrm{eff}}$ changes sign at $y=0.075$. As we will
show in the next section, this change in $J_{1}^{\mathrm{eff}}$ is
the driving force behind the change in the magnetic order that happens
in Fe$_{1+y}$Te above a critical
concentration of Fe excess.

\section{Classical phase diagram}\label{PD}

With the $y$-dependent exchange constants shown in Fig. 6 (b), we
now proceed to the solution of the effective super-exchange model
(\ref{exchange-model}). We compute the classical phase diagram by
numerical minimization of its classical energy with the constraint
that all spins have unit length. Our findings are summarized in the
classical phase diagram presented in Fig. 7. To look for a wide variety
of states, we introduce four sublattices, labeled as 0, 1, 2, and
3 (see Fig. 8) and perform full minimization of the classical energy.
To each sublattice we associate a local frame given by angles $\varphi_{0},\varphi_{1},\varphi_{2},\varphi_{3}$.
We define the global reference frame by setting $\varphi_{0}=0$.
In addition, we consider only spin configurations which can be characterized
by a single-${\bf {q}}$ spiral, such that after a translation from
site to site in the same sublattice, the magnetic moment rotates by
an angle $\theta={\bf q}\cdot{\bf r}$, where ${\bf r}=2\, m\, a_{x}+2\, n\, a_{y}$,
$m$ and $n$ are integers, ${\bf a}=(a_{x},a_{y})$ is the lattice
vector. Then, the general expression for the on-site magnetization
is given by
\begin{eqnarray}
{\bar{{\bf S}}}_{\mu}({\bf r})={\hat{{\bf x}}}\,\sin({\bf q\cdot{\bf r+\varphi_{\mu})+{\hat{{\bf y}}}\,\cos({\bf q\cdot{\bf r+\varphi_{\mu}),\label{eq:S_{b}ar}\label{eq:Sm}}}}}
\end{eqnarray}
 where $\mu=0,1,2,3$ is the sublattice index. Substituting ${\bar{{\bf S}}}_{\mu}({\bf r})$
into Eq.(\ref{exchange-model}) yields the classical energy $E_{{\rm cl}}=E(\{\varphi_{\mu}\},{\bf q)}$.
The explicit expression for the classical energy is rather cumbersome
and, for convenience, is given in Appendix B. We minimize $E_{cl}$
numerically and for each set of parameters $J_{ij}^{\mathrm{eff}}$
and $K_{ij}$ we find the classical ground state characterized by
$\varphi_{1},\,\varphi_{2},\,\varphi_{3},\, q_{x}$, and $q_{y}$.
In our computation we fix the second-neighbor biquadratic exchange
to be equal to $K_{2}=3.0$ meV/$S^2$,
 but our results do not depend strongly on this value. We vary therefore
two parameters of the model: the nearest-neighbor biquadratic exchange
$K_{1}$ and the excess iron concentration $y$,
 which affects indirectly the effective exchange couplings $J_{ij}^{\mathrm{eff}}$
via the RKKY interaction.

In Figs. 8 (a)-(c), we draw the real-space spin
 configurations presented in the phase diagram of Fig. 7. Among all
possible states, in the parameter space presented in the phase diagram,
only three states are realized: the stripe phase characterized by
$\varphi_{1}=0,\,\varphi_{2}=\pi,\,\varphi_{3}=\pi,\, q_{x}=0,\, q_{y}=0$
shown in Fig. 8 (a), the double stripe phase characterized by $\varphi_{1}=0,\,\varphi_{2}=0,\,\varphi_{3}=\pi,\, q_{x}=\pi/2,\, q_{y}=\pi/2$
shown in Fig. 8 (b) and the incommensurate spiral (labeled as IC)
state shown in Fig. 8 (c) characterized by $\varphi_{1}=\pi/2-\delta,\,\varphi_{2}=\pi-2\delta,\,\varphi_{3}=\pi/2-\delta,q_{x}=\pi/2-\delta,\, q_{y}=\pi/2-\delta$.
Note that in Fig. 8 (c) we show the IC state with an exaggerated canting
angle. To clarify the structure of these different orderings, we also
take the Fourier transforms of the ground state spin configuration
obtained by the minimization and then compute the corresponding spin
structure factor. As expected, the structure factor exhibits peaks
at the following ordering wave-vectors: ${\mathbf{Q}}=(0,\pi)$ for
the stripe phase,
 ${\mathbf{Q}}=(\pi/2,\pi/2)$ for the bicollinear double-stripe phase,
and ${\mathbf{Q}}=(\pi/2-\delta,\pi/2-\delta)$ for the IC phase.

The structure of the phase diagram (see Fig. 7) can be summarized
as follows. For small values of $y$, there is a thin strip of the
single-stripe phase. This phase is stabilized by a strong $J_{2}^{\mathrm{eff}}$
coupling (see Fig. 6 (b)) and quickly disappears because $J_{2}^{\mathrm{eff}}$
decreases rapidly with increasing $y$. We believe that the stripe
phase has not been observed in Fe$_{1+y}$Te because all known compounds
belonging to this family are, actually, nonstoichiometric and have
a small amount of interstitial Fe significantly larger than the boundary
value of $y=0.014$ below which the stripe phase is stable.

The central region in the phase diagram $(y>0.014)$ is occupied by
the bicollinear double-stripe state. This is the state which is experimentally
observed in the Fe$_{1+y}$Te compound with a low level of excess
Fe. The stability of this phase over a wide range of parameters is
explained by the smallness of the effective nearest-neighbor coupling
$J_{1}^{\mathrm{eff}}$ and relatively strong strength of the 
third-neighbor coupling $J_{3}^{\mathrm{eff}}$. The biquadratic couplings
$K_{ij}$ among which the dominating role is played by $K_{2}$ and
$K_{\mathrm{diag}}=-K_{2}$ (see Appendix A), also play an important
role in stabilizing this state: except in the region near $y=0.075$,
a finite value of the nearest-neighbor biquadratic coupling $K_{1}$
is necessary to stabilize the double stripe phase over the incommensurate
spiral state. The region near $y=0.075$ is rather peculiar, as there
the effective nearest neighbor coupling $J_{1}^{\mathrm{eff}}$ is
equal to zero or is very small compared with the other interactions,
making the bicollinear double stripe state the most stable one even
in the absence of the biquadratic exchange. The rest of the phase
diagram is occupied by an IC phase (see, Fig. 8 (c)), which is the
$(q,q)$ spiral state experimentally observed for sufficiently large
$y$.

\begin{figure}
\label{fig7} \includegraphics[width=0.99\columnwidth]{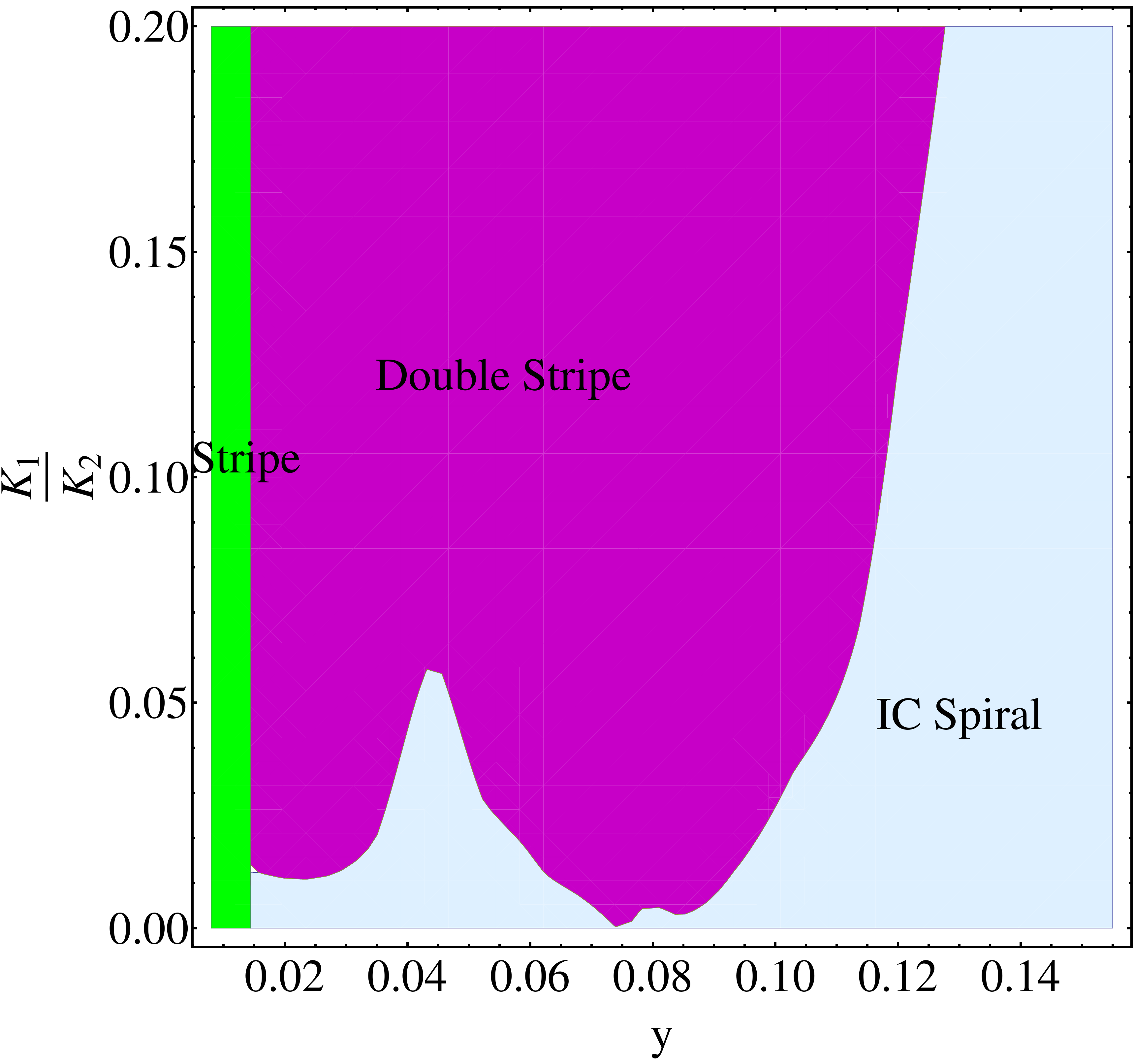} \caption{(Color online) Magnetic phase diagram of the effective low-energy model (\ref{exchange-model})
as a function of Fe excess $y$ and the ratio between the first- and
second-neighbor biquadratic exchanges, $\frac{K_{1}}{K_{2}}$, computed
with (in meV/$S^2$ units) $J_{1}^{\mathrm{eff}}(y)=-3.4+J_{1}^{\mathrm{RKKY}}(y)$,
$J_{2}^{\mathrm{eff}}(y)=11.6+J_{2}^{\mathrm{RKKY}}(y)$, $J_{3}^{\mathrm{eff}}(y)=15.1+J_{3}^{\mathrm{RKKY}}(y)$.
 The RKKY interactions $J_{ij}^{\mathrm{RKKY}}$ are shown in Fig.
\ref{fig6}. We set $K_{2}=3$. }
\end{figure}

\begin{figure*}
\label{fig8} \includegraphics[width=0.65\columnwidth]{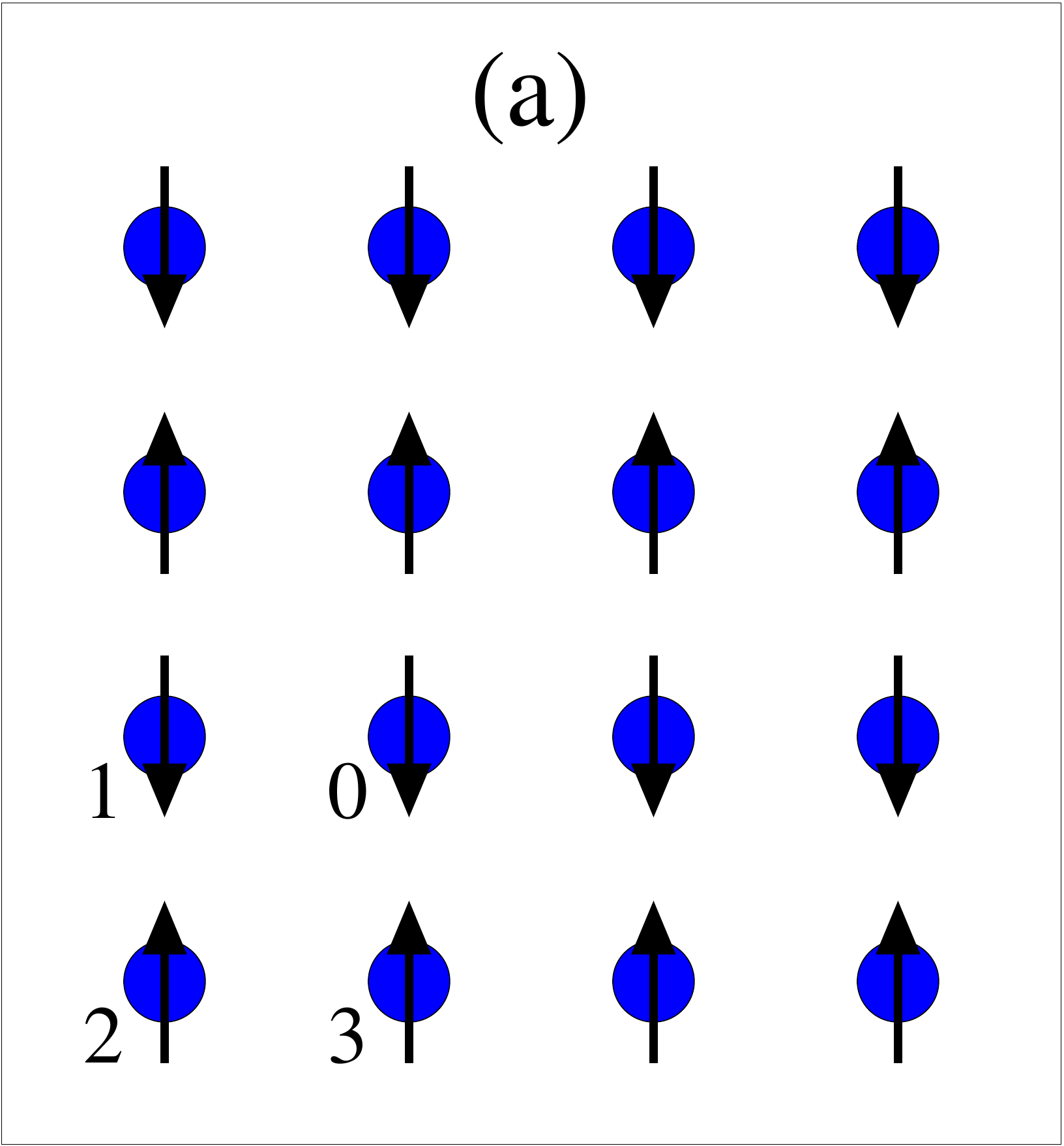} \includegraphics[width=0.65\columnwidth]{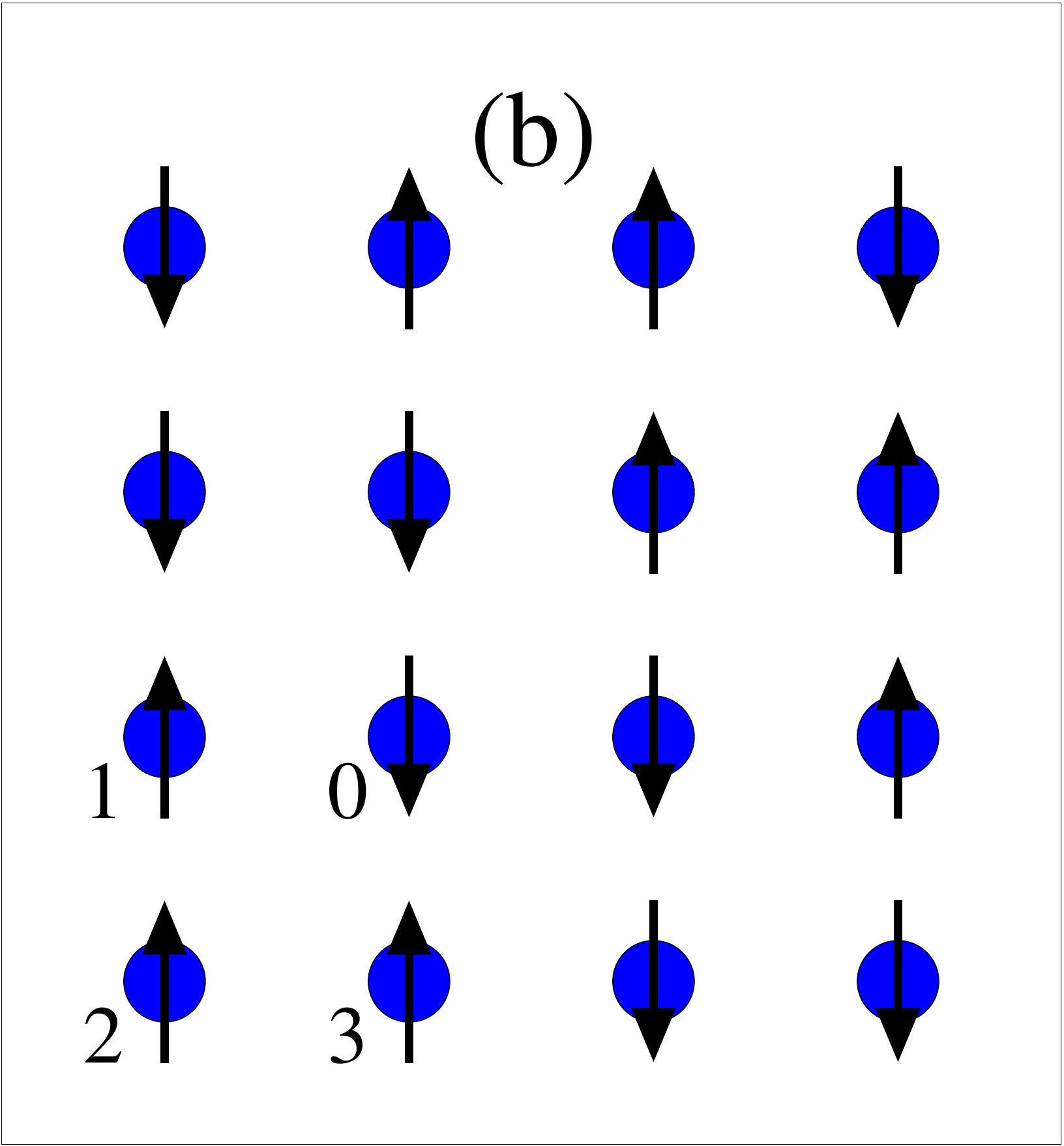}
\includegraphics[width=0.65\columnwidth]{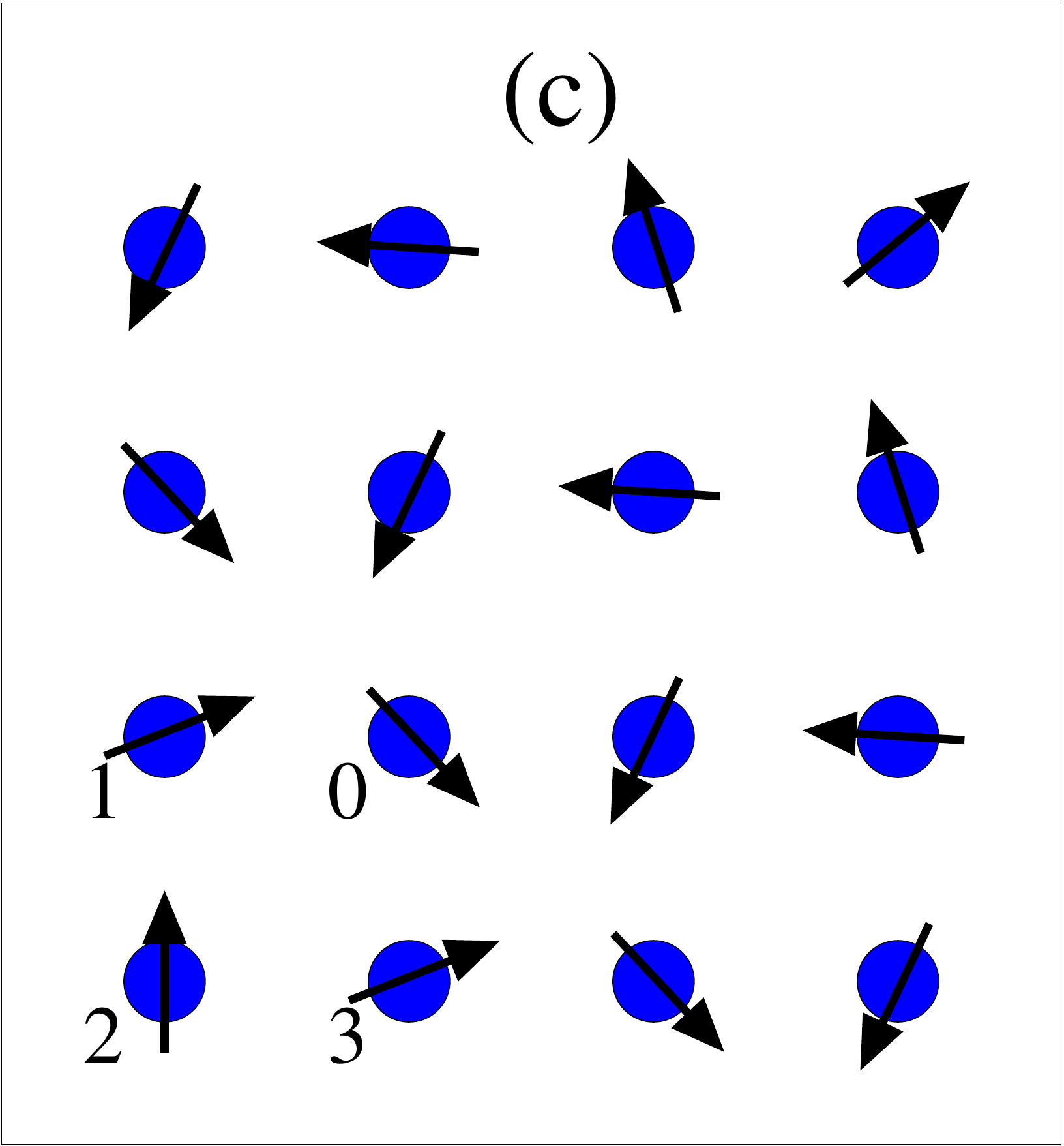} \includegraphics[width=0.65\columnwidth]{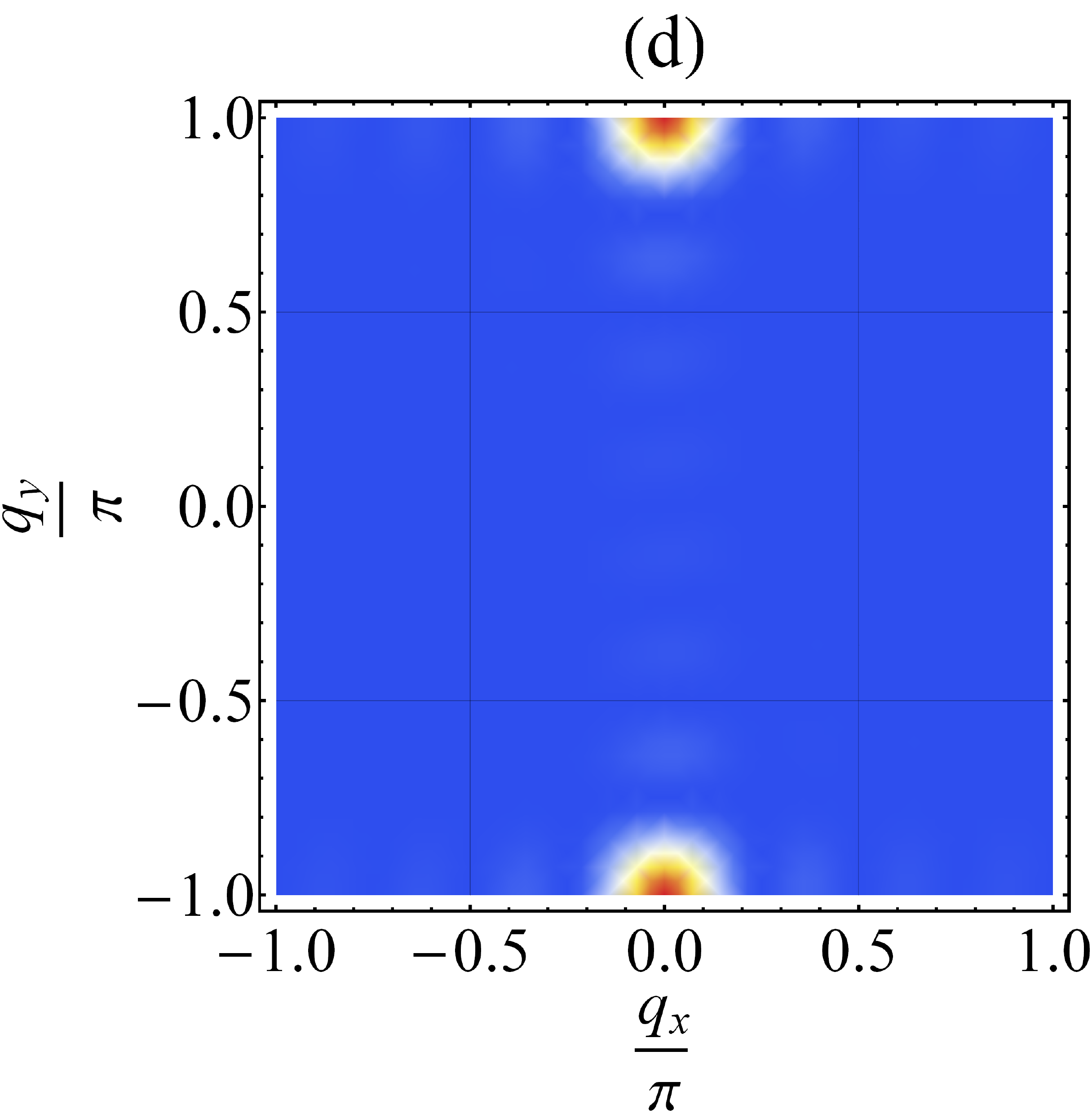}
\includegraphics[width=0.65\columnwidth]{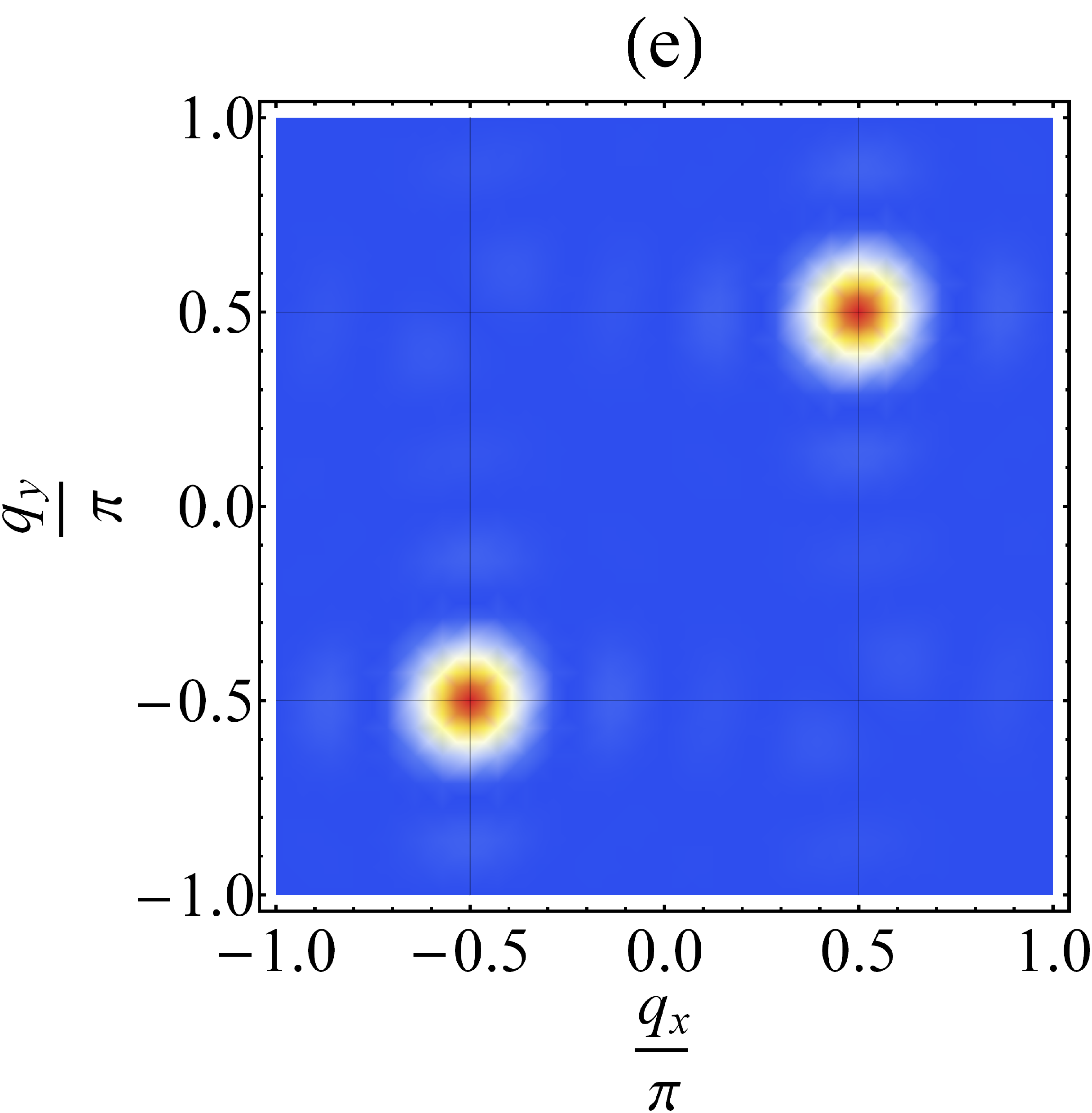} \includegraphics[width=0.65\columnwidth]{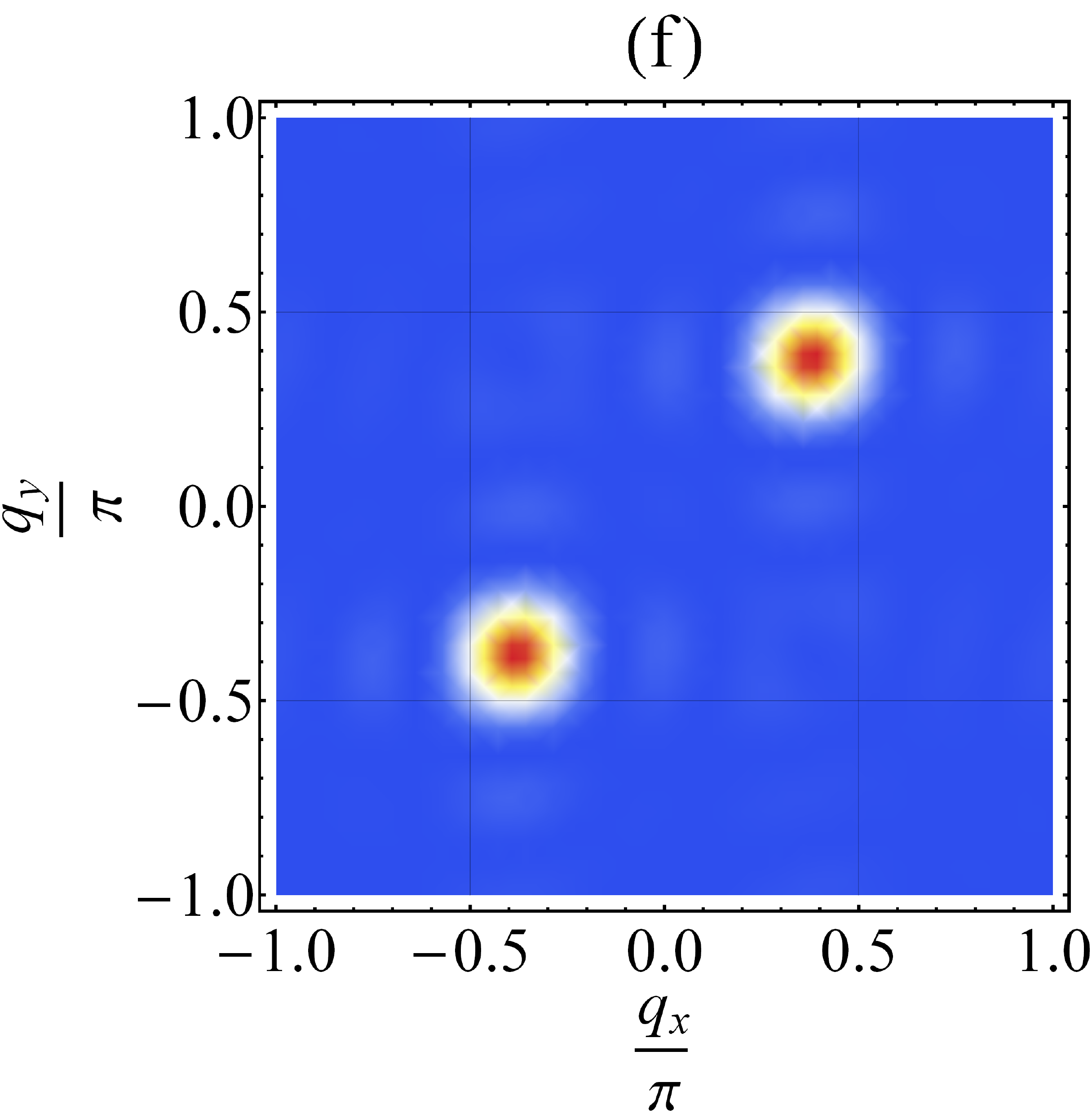}
\caption{(Color online) Schematic representations of the spin
 configurations in the ground state obtained by the minimization of
the classical energy with respect to $\varphi_{1},\varphi_{2},\varphi_{3},q_{x}$,
and $q_{y}$: (a) $\varphi_{1}=0,\varphi_{2}=\pi,\varphi_{3}=\pi,q_{x}=0,q_{y}=0$
gives the single-stripe phase, (b) $\varphi_{1}=0,\varphi_{2}=0,\varphi_{3}=\pi,q_{x}=\pi/2,q_{y}=\pi/2$
gives the double-stripe phase, (c) $\varphi_{1}=\pi/2-\delta,\varphi_{2}=\pi-2\delta,\varphi_{3}=\pi/2-\delta,q_{x}=\pi/2-\delta,q_{y}=\pi/2-\delta$
 gives the incommensurate spiral phase. (d),(e) The structure factors
computed for the magnetic orders displayed in (a)-(c), respectively.
Bright spots correspond to the sharp peaks that appear at the corresponding
ordering wavevectors. }
\end{figure*}

In the experiment performed in Ref. \cite{koz13}, the phase transition
from the bicollinear double stripe to the IC phase was observed approximately
for $y\approx0.11$. This is consistent with our phase diagram presented
in Fig. 7 if one takes $\frac{K_{1}}{K_{2}}\simeq0.1$, which seems
to be a realistic ratio, since the magneto-elastic coupling favors
the biquadratic exchange between second neighbors (see Appendix A
for more details). We emphasize that, because our model contains assumptions
about the bare values of $J_{ij}$ and the amount of electrons introduced
by each excess Fe, the precise value of $y$ for which the transition
takes place is beyond our scope. Yet, the general tendency of a double-stripe
to IC transition for increasing charge doping encoded in the phase
diagram of Fig. 7 is
 robust and consistent with the experimental observations.

\section{Conclusions}
\label{Dis}

In summary, we have studied the evolution of the magnetic order in
Fe$_{1+y}$Te
 as function of $y$. Starting with a model containing both localized
spins and itinerant electrons, we derived an effective superexchange
Hamiltonian to describe the magnetic properties of Fe$_{1+y}$Te which
contains both the long-range RKKY-type spin-spin interaction mediated
by the itinerant electrons and the biquadratic interactions due to
magneto-elastic
effects. Thus, $y$-dependent exchange interactions naturally arise
in our model due to changes in the low-energy itinerant electronic
states promoted by charge doping.

After calculating the classical phase diagram of the $y$-dependent
effective superexchange model, we showed that Fe$_{1+y}$Te has a
general tendency for a double-stripe to incommensurate-spiral transition
with increasing excess iron
concentration. In particular, for small $y$, the magnetic order is
a double-stripe state, arising due to the presence of a significant
antiferromagnetic $J_{3}^{\mathrm{eff}}$ coupling, which has mostly
localized origin. Beyond a certain critical value of $y$, the incommensurate
spiral state becomes the most stable. This transition is driven not
only by the suppression of the antiferromagnetic exchanges $J_{3}^{\mathrm{eff}}$
and $J_{2}^{\mathrm{eff}}$, but also by the enhancement (in absolute
value) and sign change of the nearest-neighbor $J_{1}^{\mathrm{eff}}$
interaction, which changes from antiferromagnetic to ferromagnetic
with increasing $y$. These changes are caused by the $y$-dependent
RKKY part of the interaction, and ultimately can be attributed to
the suppression of the $\left(\pi,0\right)/\left(0,\pi\right)$ peaks
in the itinerant spin-susceptibility and the transfer of magnetic
spectral weight from the vicinities of $\left(\pi,\pi\right)$ to
the vicinities of $\left(0,0\right)$. These changes, in turn, are
a direct consequence of the changes in the Fermi surface of the itinerant
electrons caused by the charge doping introduced by the excess Fe.
Experimental data showing the sign of $J_{1}^{\mathrm{eff}}$ to be
dependent on the interstitial Fe concentration would be a strong validation
of our model. The physics of Fe$_{1+y}$Te discussed here bares many
similarities with extensively studied double-exchange magnets, and
in particular with manganites, whose effective coupling constants
were shown to be significantly modified by charge doping.~\cite{perkins99,mancini01,jackeli02}
In both cases, the interplay between local moments and itinerant electrons
leads to a rich behavior and to the appearance of new ground states
in the classical phase diagram.

\textit{Acknowledgement.} We acknowledge useful conversations with A. Chubukov,
I. Eremin, M. Gingras, M. Imada, D.H. Lee, I. Mazin, I. Paul, S. and U. Roessler and I.Zaliznyak.
We especially thank F. Wang and Z.-Y. Lu for providing us with the unfolded five orbital tight-binding model, which we use in this work.
N.B.P. and S. D. are supported by NSF grant DMR-1255544. N.B.P. and
R. M. F. acknowledge the hospitality of the Aspen Center for Physics,
and also NSF grant No.1066293 supporting the center.

\appendix

\begin{figure*}
\label{figA1} \includegraphics[width=0.72\columnwidth]{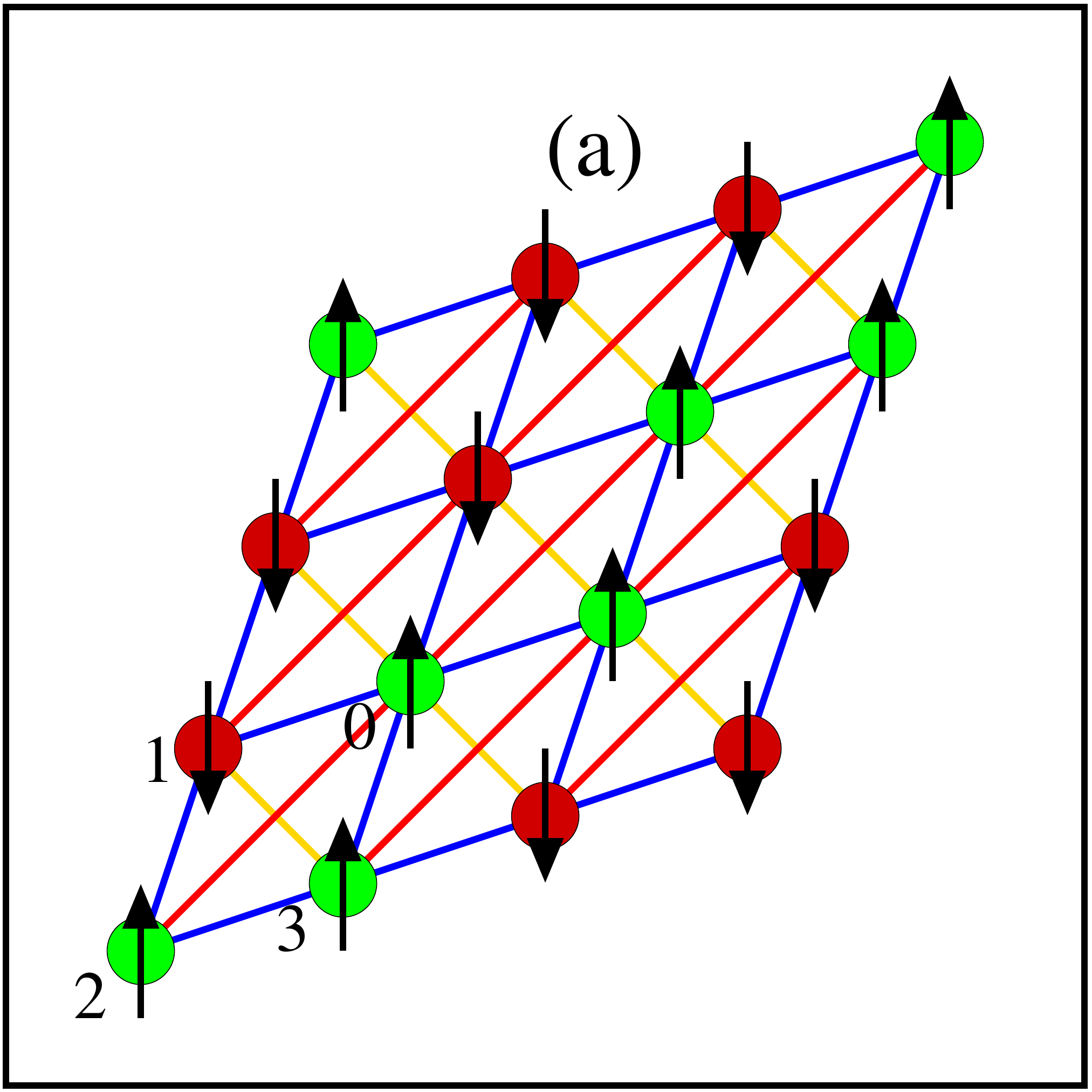} \includegraphics[width=0.62\columnwidth]{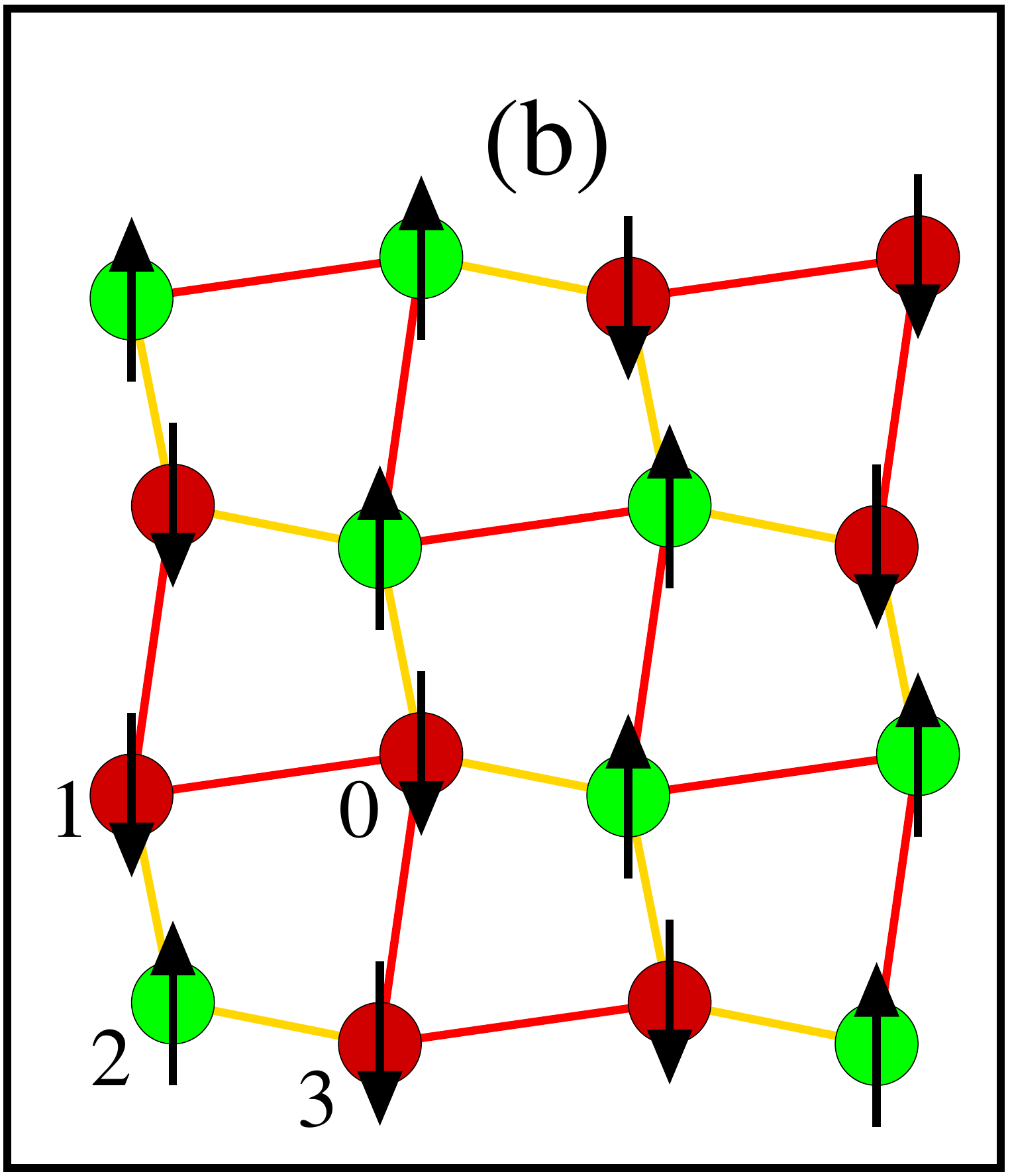}
\includegraphics[width=0.62\columnwidth]{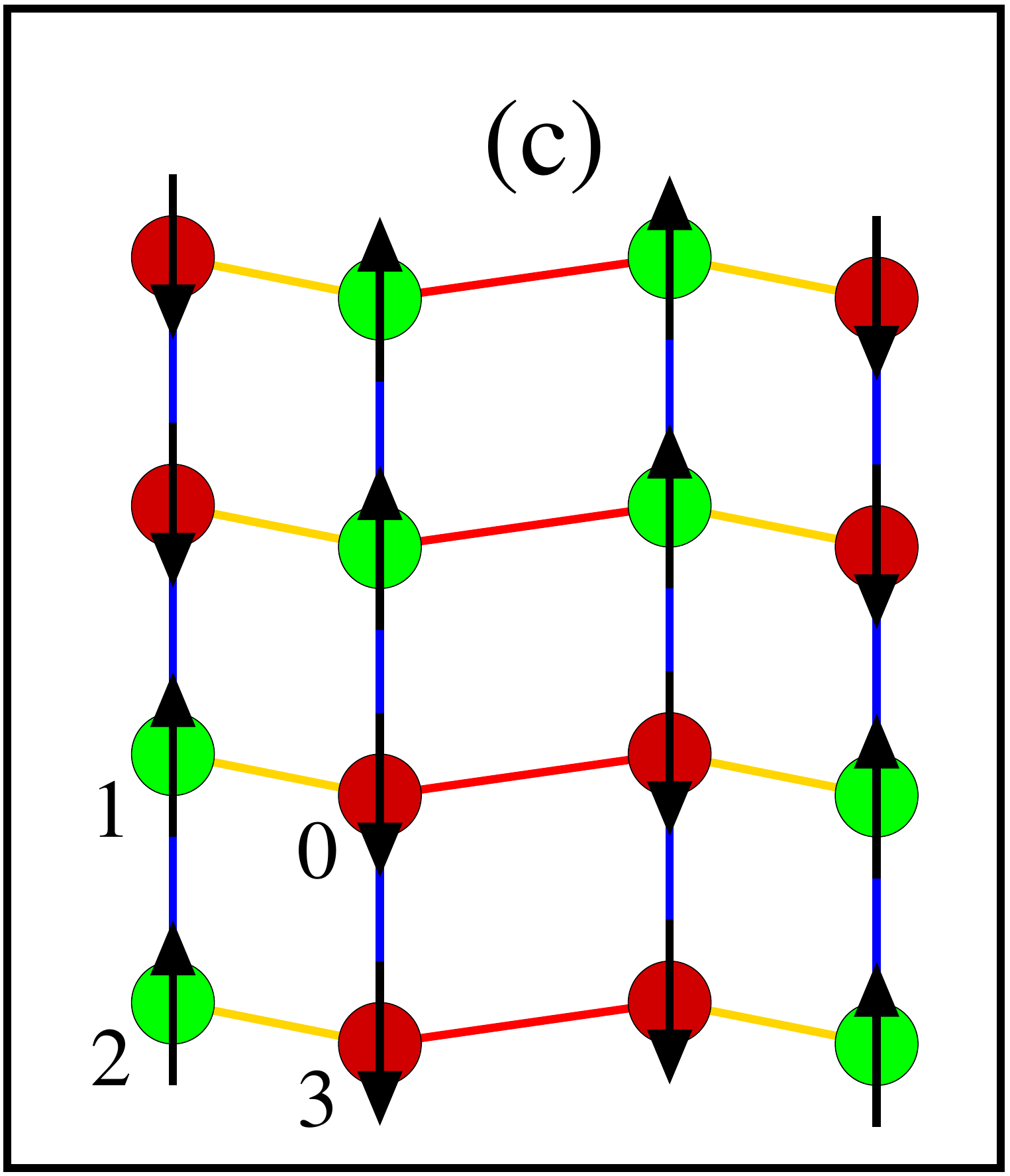} \caption{Most relevant elastic modes in Fe$_{1+y}$Te. (a) The uniform monoclinic
mode, $u_{xy}$. (b) The non-uniform mode corresponding to ${u}_{5}^{x}={u}_{5}^{y}$.
(c) The non-uniform mode corresponding to the ${u}_{6}$ distortion.
The mode corresponding to the ${u}_{7}$ distortion would display
the same configuration but rotated by 90$^\circ$. We use the following
convention: red bonds are lengthened with respect to the tetragonal
lattice, yellow bonds are shortened, blue bonds remain of the same
length. Green (red) sites have spins ferromagnetically (antiferromagnetically)
aligned with each other. }
\end{figure*}

\section{Derivation of biquadratic exchange couplings}

\label{App}

There are several microscopic mechanisms which lead to non-Heisenberg
exchange couplings such as, e.g., biquadratic and ring exchanges.
Among all of them, one of the most effective ways to induce a fairly
strong non-Heisenberg exchange is through the coupling to the lattice
via the magneto-elastic effect. In the pnictides and chalcogenides,
this key role played by the magnetoelastic coupling was extensively
discussed, see Refs. \cite{Fernandes10,paul11,Paul11_PRL}.

Here, we present a brief discussion of the possible magnetoelastic
origin of the first- and the second-neighbor biquadratic exchange
interactions $K_{1}$ and $K_{2}$ in Fe$_{1+y}$Te. Motivated by
the experimentally measured lattice distortions, which are small,
we assume a linear regime in which exchange interactions and elastic
energies depend only on the distance between lattice sites. As the
magneto-elastic Hamiltonian involves more than one normal mode of
a square lattice, it is convenient to follow the notation of Ref.
\cite{paul11} and describe the lattice distortions by the strain
tensor separated in uniform and non-uniform parts:
\begin{eqnarray}
u_{ij}({\bf r})=u_{ij}+\frac{\imath}{2}\sum_{q\neq0}(q_{i}u_{j}({\bf q})+q_{i}u_{j}({\bf q}))e^{\imath{\bf q}\cdot{\bf r}}.
\end{eqnarray}

The most relevant $\mathbf{q}=0$ lattice modulation in Fe$_{1+y}$Te
is the monoclinic distortion, given by $u_{xy}\equiv\partial_{y}u_{x}+\partial_{x}u_{y}$,
and illustrated in Fig. 9(a). It corresponds to a distortion of the
square in a rhombus with a short and a long diagonal. The most relevant
non-uniform lattice modulations are those with modulation vectors
${\bf q}_{5}=(\pi,\pi)$, ${\bf q}_{6}=(\pi,0)$ and ${\bf q}_{7}=(0,\pi)$.
To make the notations simpler, we denote $u({\bf q}_{5})\equiv u_{5}$,
$u({\bf q}_{6})\equiv u_{6}$ and $u({\bf q}_{7})\equiv u_{7}$. The
real space patterns of the first two are shown in Figs. 9(b)-(c).
The non-uniform mode corresponding to ${u}_{5}$ generates a distortion
with ladders along the diagonal of short and long nearest-neighbor
bonds. On the other hand, the non-uniform mode corresponding to the
${u}_{6}$ ($u_{7}$) distortion has bonds that alternate between
short and long in the $x$-direction ($y$-direction).

Using this notation, the dominant magneto-elastic term is given by
\cite{paul11}
\begin{eqnarray}
H_{\text{ME}} & = & g_{1}(\mathbf{S}_{1}\cdot\mathbf{S}_{3}-\mathbf{S}_{0}\cdot\mathbf{S}_{2})u_{xy}\label{ME}\\
 & + & g_{2}\left[(\mathbf{S}_{2}\cdot\mathbf{S}_{3}-\mathbf{S}_{0}\cdot\mathbf{S}_{1}){u}_{5}^{x}+(\mathbf{S}_{1}\cdot\mathbf{S}_{2}-\mathbf{S}_{0}\cdot\mathbf{S}_{3}){u}_{5}^{y}\right]\nonumber \\
 & + & g_{3}\left[(\mathbf{S}_{2}\cdot\mathbf{S}_{3}+\mathbf{S}_{0}\cdot\mathbf{S}_{1}){u}_{6}^{x}+(\mathbf{S}_{1}\cdot\mathbf{S}_{2}+\mathbf{S}_{0}\cdot\mathbf{S}_{3}){u}_{7}^{y}\right].\nonumber
\end{eqnarray}
 where $g_{i}$ are the magneto-elastic couplings. The spins $\mathbf{S}_{i}$
with $i=0,...,3$ correspond to the four spins in the sublattice shown
in Fig. 8a of the main text.

Because the elastic modes are assumed to be non-critical, the elastic
energy acquires a harmonic form:
\begin{eqnarray}
H_{\text{elast}}=\frac{c_{66}}{2}u_{{xy}}^{2}+\frac{{\Omega}_{1}}{2}{u}_{5}^{2}+\frac{{\Omega}_{2}}{2}({u}_{6}^{2}+{u}_{7}^{2})~,\label{Elastic}
\end{eqnarray}
 where the constants $c_{66}$, $\Omega_{1}$, and $\Omega_{2}$ represent
the elastic stiffness of the different lattice distortions described
by $u_{{xy}}$, ${u}_{5}$, and ${u}_{6}/u_{7}$, respectively. The
equilibrium lattice distortions $u_{xy}$, $u_{5}$, $u_{6}$, and
$u_{7}$ are found by minimizing the energy:
\begin{eqnarray}
u_{xy} & = & -\frac{g_{1}}{c_{66}}(\mathbf{S}_{1}\cdot\mathbf{S}_{3}-\mathbf{S}_{0}\cdot\mathbf{S}_{2})\nonumber \\
u_{5}^{x} & = & -\frac{g_{2}}{{\Omega}_{1}}(\mathbf{S}_{2}\cdot\mathbf{S}_{3}-\mathbf{S}_{0}\cdot\mathbf{S}_{1})\nonumber \\
u_{5}^{y} & = & -\frac{g_{2}}{{\Omega}_{1}}(\mathbf{S}_{1}\cdot\mathbf{S}_{2}-\mathbf{S}_{0}\cdot\mathbf{S}_{3})\\
u_{6} & = & -\frac{g_{3}}{{\Omega}_{2}}(\mathbf{S}_{2}\cdot\mathbf{S}_{3}+\mathbf{S}_{0}\cdot\mathbf{S}_{1})\nonumber \\
u_{7} & = & -\frac{g_{3}}{{\Omega}_{2}}(\mathbf{S}_{1}\cdot\mathbf{S}_{2}+\mathbf{S}_{0}\cdot\mathbf{S}_{3}).\nonumber
\end{eqnarray}

Integrating out the lattice distortions, we obtain the following biquadratic
Hamiltonian:
\begin{eqnarray}
H_{{\rm {bi}}} & = & -\frac{g_{1}^{2}}{2c_{66}}(\mathbf{S}_{1}\cdot\mathbf{S}_{3}-\mathbf{S}_{0}\cdot\mathbf{S}_{2})^{2}\label{ME Corrections}\\
 & - & \frac{g_{2}^{2}}{2\mathbf{\Omega}_{1}}[(\mathbf{S}_{2}\cdot\mathbf{S}_{3}-\mathbf{S}_{0}\cdot\mathbf{S}_{1})^{2}+(\mathbf{S}_{1}\cdot\mathbf{S}_{2}-\mathbf{S}_{0}\cdot\mathbf{S}_{3})^{2}]\nonumber \\
 & - & \frac{g_{3}^{2}}{2\mathbf{\Omega}_{2}}[(\mathbf{S}_{2}\cdot\mathbf{S}_{3}+\mathbf{S}_{0}\cdot\mathbf{S}_{1})^{2}+(\mathbf{S}_{1}\cdot\mathbf{S}_{2}+\mathbf{S}_{0}\cdot\mathbf{S}_{3})^{2}].\nonumber
\end{eqnarray}

Here it is convenient to rewrite this expression explicitly in terms
of the first- and second-neighbor biquadratic interactions $K_{1}$
and $K_{2}$, as well as of the ring exchange interactions $K_{\square}$
and $K_{{\rm diag}}$:
\begin{widetext}
\begin{eqnarray}
H_{{\rm {bi}}}= &  & -\left(\frac{g_{2}^{2}}{2\mathbf{\Omega}_{1}}+\frac{g_{3}^{2}}{2\mathbf{\Omega}_{2}}\right)[(\mathbf{S}_{2}\cdot\mathbf{S}_{3})^{2}+(\mathbf{S}_{0}\cdot\mathbf{S}_{1})^{2}+(\mathbf{S}_{1}\cdot\mathbf{S}_{2})^{2}+(\mathbf{S}_{0}\cdot\mathbf{S}_{3})^{2}]-\frac{g_{1}^{2}}{2c_{66}}[(\mathbf{S}_{1}\cdot\mathbf{S}_{3})^{2}+(\mathbf{S}_{0}\cdot\mathbf{S}_{2})^{2}]\\
 &  & +\left(\frac{g_{2}^{2}}{\mathbf{\Omega}_{1}}-\frac{g_{3}^{2}}{\mathbf{\Omega}_{2}}\right)[(\mathbf{S}_{2}\cdot\mathbf{S}_{3})(\mathbf{S}_{0}\cdot\mathbf{S}_{1})+(\mathbf{S}_{1}\cdot\mathbf{S}_{2})(\mathbf{S}_{0}\cdot\mathbf{S}_{3})]+\frac{g_{1}^{2}}{c_{66}}(\mathbf{S}_{1}\cdot\mathbf{S}_{3})(\mathbf{S}_{0}\cdot\mathbf{S}_{2})\nonumber \\
= &  & -K_{1}\sum_{\langle ij\rangle}\left(\mathbf{S}_{i}\cdot\mathbf{S}_{j}\right)^{2}-K_{2}\sum_{\langle\langle ij\rangle\rangle}\left(\mathbf{S}_{i}\cdot\mathbf{S}_{j}\right)^{2}-K_{\square}\sum_{\square}(\mathbf{S}_{i}\cdot\mathbf{S}_{j})(\mathbf{S}_{k}\cdot\mathbf{S}_{l})-K_{{\rm diag}}\sum_{\langle\langle ij\rangle\rangle}(\mathbf{S}_{1}\cdot\mathbf{S}_{3})(\mathbf{S}_{0}\cdot\mathbf{S}_{2})~,\nonumber
\end{eqnarray}
 \end{widetext} where
\begin{eqnarray*}
K_{1} & = & \frac{g_{2}^{2}}{{\Omega}_{1}}+\frac{g_{3}^{2}}{{\Omega}_{2}}\\
K_{2} & = & \frac{g_{1}^{2}}{c_{66}}\\
K_{\square} & = & -\frac{g_{2}^{2}}{{\Omega}_{1}}+\frac{g_{3}^{2}}{{\Omega}_{2}}\\
K_{{\rm diag}} & = & -\frac{g_{1}^{2}}{c_{66}}
\end{eqnarray*}

Because we expect the non-uniform strains ${u}_{5}$, ${u}_{6}$,
and ${u}_{7}$ to have a stronger stiffness than the stiffness of
the monoclinic distortion $u_{xy}$, in our calculations we neglected
the ring exchange term $K_{\square}$ and assumed that $K_{1}\ll K_{2}$.
Furthermore, we used the fact that $K_{\mathrm{diag}}=-K_{2}$.

\section{Classical energy}

\begin{widetext}

Here we present the expression for the classical energy of the local-spin
model of Eq. (\ref{exchange-model}), as function of the local angles
$\varphi_{i}$ and the ordering vector $\left(q_{x},q_{y}\right)$:

\begin{eqnarray*}
E_{cl} & = & \frac{S^{2}}{4}\Big[J_{1}\Big(\cos\varphi_{1}+\cos\left(\varphi_{1}+2q_{x}\right)+\cos\left(\varphi_{3}-\varphi_{2}\right)+\cos\left(\varphi_{3}-(\varphi_{2}+2q_{x})\right)+\cos\varphi_{3}+\cos\left(\varphi_{3}+2q_{y}\right)\\
 & + & \cos\left(\varphi_{1}-\varphi_{2}\right)+\cos\left(\varphi_{1}-(\varphi_{2}+2q_{y})\right)\Big)+J_{2}\Big(\cos\varphi_{2}+\cos\left(\varphi_{2}+2q_{x}+2q_{y}\right)+\cos\left(\varphi_{1}-(\varphi_{3}+2q_{y})\right)\\
 & + & \cos\left(\varphi_{1}+2q_{x}-\varphi_{3}\right)+\cos\left(\varphi_{2}+2q_{y}\right)+\cos\left(\varphi_{1}-\varphi_{3}\right)+\cos\left(\varphi_{2}+2q_{x}\right)+\cos\left((\varphi_{1}+2q_{x})-(\varphi_{3}+2q_{y})\right)\Big)\\
 & + & 4J_{3}\Big(\cos2q_{x}+\cos2q_{y}\Big)+K_{1}S^{2}\Big(\cos^{2}\varphi_{1}+\cos^{2}\left(\varphi_{1}+2q_{x}\right)+\cos^{2}\left(\varphi_{3}-\varphi_{2}\right)+\cos^{2}\left(\varphi_{3}-(\varphi_{2}+2q_{x})\right)\\
 & + & \cos^{2}\varphi_{3}+\cos^{2}\left(\varphi_{3}+2q_{y}\right)+\cos^{2}\left(\varphi_{1}-\varphi_{2}\right)+\cos^{2}\left(\varphi_{1}-(\varphi_{2}+2q_{y})\right)\Big)+K_{2}S^{2}\Big(\cos^{2}\varphi_{2}+\cos^{2}\left(\varphi_{2}+2q_{x}+2q_{y}\right)\\
 & + & \cos^{2}\left(\varphi_{1}-(\varphi_{3}+2q_{y})\right)+\cos^{2}\left(\varphi_{1}+2q_{x}-\varphi_{3}\right)+\cos^{2}\left(\varphi_{2}+2q_{y}\right)+\cos^{2}\left(\varphi_{1}-\varphi_{3}\right)+\cos^{2}\left(\varphi_{2}+2q_{x}\right)\\
 & + & \cos^{2}\left((\varphi_{1}+2q_{x})-(\varphi_{3}+2q_{y})\right)+\cos\varphi_{2}\cos\left(\varphi_{1}-\varphi_{3}\right)+\cos\left(\varphi_{2}+2q_{x}\right)\cos\left(\varphi_{1}+2q_{x}-\varphi_{3}\right)\\
 & + & \cos\left(\varphi_{2}+2q_{y}\right)\cos\left(\varphi_{1}-\varphi_{3}-2q_{y}\right)+\cos\left(\varphi_{2}+2q_{x}+2q_{y}\right)\cos\left(\varphi_{1}+2q_{x}-\varphi_{3}-2q_{y}\right)\Big)\Big]~.
\end{eqnarray*}
 \end{widetext}

\pagebreak{}
\end{document}